\documentstyle[amssymb,prb,aps,multicol,epsfig]{revtex} 
\begin{document}

\title{Self generated  randomness, defect wandering and viscous flow in stripe glasses}  
\author{Harry Westfahl Jr.$^{(a)}$, J\"org Schmalian$^{(a)}$, and Peter G Wolynes$^{(b)}$.}  
\address{$^{(a)}$Department of Physics and Astronomy and Ames Laboratory,
Iowa State University, Ames, IA 50011\\
$^{(b)}$ Department of Chemistry and Biochemistry, University of California, San Diego, La Jolla, CA 92093 }  
\date{\today}  
\maketitle  
\vskip 0.5cm  
\begin{abstract}  
\leftskip 54.8pt \rightskip 54.8pt We show that the competition between interactions on different length  
scales, as relevant for the formation of stripes in doped Mott insulators, 
can cause a glass transition in a system with no explicitly quenched disorder.
We analytically determine a universal criterion for the emergence of an exponentially large number  
of metastable configurations that leads to a finite configurational 
entropy and a landscape dominated viscous flow. We demonstrate that glassines 
is unambiguously tied to a new length scale which characterizes the typical length over which defects
and imperfections in the stripe pattern are allowed to wander over long
times. 
\end{abstract}  
\vskip 0.5cm

 
\begin{multicols}{2}  
\narrowtext   
 
\section{Introduction}

The competition of interactions on different length scales is one of the
mechanisms able to stabilize mesoscale phase separations and create spatial
inhomogeneities in a wide variety of systems. The most typical situation is
a competition between short-ranged forces, that favors the formation of a
uniform condensed phase, and the long-range forces which can energetically
frustrate this condensation. Classical examples are the formation of domains
in magnetic multilayer compounds\cite{GD82,AB92}, mesoscopic structures
built by assembling polymers in solution or amphiphiles in\ water-oil
mixtures \cite{GT83,GBS96}.Very often, these systems exhibit a long time
dynamics similar to the relaxation seen in glasses.\smallskip\ Conversely,
many proposals have been made that the glassy behavior of molecular liquids
might arise from frustration of specific crystalline-like orders
incompatible with global packing, e.g., icosahedral order in dense liquids 
\cite{DRN85}.

The observation of complex orbital and charge patterns in CMR-manganites or
of charge stripes in doped nickelates and cuprates\cite{CCJ93,JTra95}
suggests that a similar competition causes inhomogeneous structures in these
strongly correlated electron systems\cite{EK93}. This point of view is
supported by the observation of spatial \ inhomogeneities\cite
{CBJ92,LC97,TUI99,JBC99,HS99,CH99,HSS99} as well as slow, activated glassy
dynamics\cite{JBC99,CH99}, as seen in recent NMR experiments. \ Upon
cooling, the Cu-NMR signal in $La_{2-x}Sr_{x}CuO_4$ based systems disappears
(is ``wiped out''). This has been interpreted in terms of an electronic
relaxation slower than the Larmor precession of the nuclear spins\cite
{JBC99,CH99}. As a result, the NMR signal decays so fast that it simply
cannot be detected anymore. Thus the ``wipe out effect'' discussed in Refs. 
\cite{CBJ92,LC97,TUI99,JBC99,HS99,CH99,HSS99} is clear evidence for a
dramatic increase of the relaxation times of the electronic system.
Similarly, La-NMR was used to directly show that there is a glassy activated
dynamics with a maximum of $T_{1}^{-1}\left( T\right) $, separating
relaxational dynamics which is slower than the nuclear Lamor frequency at
low temperatures from faster processes at higher $T$\cite{JBC99,CH99}. The
typical activation energies of these dynamical processes have been analyzed
by Curro et al.\cite{CH99} who made the surprising observation that they are
rather independent of the specific details of added impurities etc. Also,
the width of the distribution of activation energies is comparable to its
mean value in systems with rather different chemical composition. In view of
this striking universality of the anomalous long time relaxation in doped
Mott insulators, we recently suggested that glassiness in these systems is 
{\em self generated}, i.e.,{\em \ }it does not rely on the presence of
quenched disorder\cite{SW00,SWW00}. The latter may, in general, further
stabilize a glassy state. \ This is supported by molecular dynamics
calculations for charge ordering in transition metal oxides, which found an
anomalous long time relaxation with a power spectrum similar to $1/f$-noise 
\cite{SYC98}. In addition, recently Markiewicz {\em et al.}\cite{MCPC00}{\em %
\ }analyzed neutron diffraction\cite{WSE99}, NQR\cite{CCR00}, $\mu $SR\cite
{NBB98}, anelastic relaxation\cite{CCR00} and susceptibility measurements 
\cite{WUE00}, spanning altogether more than 10 orders of magnitude of
frequency, and also find a ''universal'' behavior of the activation energies
in underdoped cuprates. They also find a good description of the
relaxational dynamics using a Vogel-Fulcher law which we predicted based on
a entropic droplet argument\cite{SW00}.

In Ref.\cite{SW00,SWW00} we showed that the competition between interactions
on different length scales causes the emergence of an exponentially large
number of metastable states (with the system size). This is generally
considered as a condition for the anomalous dynamical features of
glassiness, like aging, memory effects and ergodicity breaking. It is also
the heart of the {\em random first order transition} scenario\cite{KTW89}
for vitrification of molecular liquids, originally motivated by the
similarities between density functional theories of aperiodic crystals\cite
{SW84} and the mean field theories for random spin-glasses. This scenario is
now believed to apply to a much more general class of systems. Our result
was obtained using a new replica approach,\cite{Mon95,MP991} and by solving
the resulting many body problem numerically within the self consistent
screening approximation. In this paper we develop an analytical approach
which enables us to identify the underlying physical mechanism for
glassiness in a uniformly frustrated system. We furthermore discuss that our
results can also be obtained within a dynamical approach, where glassiness
is associated with an unconventional long time limit of the charge
correlation function.

In the next section we introduce the model we investigate and summarize the
main results of this paper. The details of our approach are presented in
section IV, subsequent to our summary of the aspects of the stripe liquid
state that will be important for our results in section III. In section V we
conclude and give a list of further open questions.

\section{Model and Overview}

This paper develops an analytical approach to glassiness in a uniformly
frustrated system. By uniformly frustrated we mean that there is a
competition of interactions on different length scales and there are no
explicitly quenched degrees of freedom, like the ones caused by additional
defects or imperfections. We study a model with local tendency towards phase
separation, frustrated by a long range interaction, which, as we will show,
has all necessary features to exhibit a glass transition and yet is simple
enough to be treated analytically. In the context of cuprate systems the
model has been proposed by Emery and Kivelson\cite{EK93} and is defined by
the Hamiltonian: 
\begin{eqnarray}
{\cal H} &=&\frac{1}{2}\int d^{d}x\left\{ r_{0}\varphi ({\bf x)}^{2}+\left(
\nabla \varphi ({\bf x)}\right) ^{2}+\frac{u}{2}\varphi ({\bf x)}^{4}\right\}
\nonumber \\
&&+\frac{Q}{8\pi }\int d^{d}x\int d^{d}x^{\prime }\frac{\varphi ({\bf x)}%
\varphi ({\bf x}^{\prime })}{\left| {\bf x-x}^{\prime }\right| }.
\label{ham11}
\end{eqnarray}
Here, $\varphi ({\bf x})$ characterizes charge degrees of freedom, with $%
\varphi ({\bf x})>0$ in \ a hole-rich region, $\varphi ({\bf x})<0$ in a
hole poor region, and $\varphi ({\bf x})=0$ if the local density equals the
averaged one. If \ $r_{0}<0$ the system tends to phase separate since we
have to guarantee charge neutrality$\left\langle \varphi \right\rangle =0$.
The coupling constant, $Q$, is a measure for the frustration between this
short range coupling and the long range Coulomb interaction. For $Q=0$ and $%
r_{0}<0$ we expect at low temperatures long range charge ordering. \ The
ordering temperature can be estimated within mean field theory as $T_{c}^{0}=%
\frac{2\pi ^{2}\left| r_{0}\right| }{u\Lambda }$, with momentum cut off $%
\Lambda $ of the order of an inverse lattice constant. As shown in Ref. \cite
{NRK99}, within a large $N$ approach (where $\varphi ({\bf x)}$ is
generalized to an $N$-component field), for all $Q>0$ , the Coulomb
interaction suppresses this ordered state at finite $T$. Instead, as
revealed by a mean field analysis of Eq.\ref{ham11} where $N\rightarrow
\infty $, the system undergoes several crossovers. 

\begin{figure}
\centerline{\epsfig{file=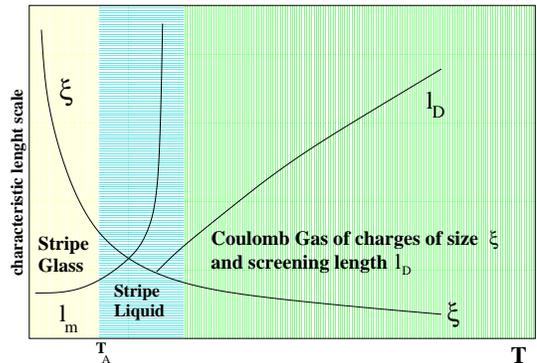,width=7cm,scale=0.85}} 
\caption{The competition between different lenght scales according to the
mean field sollution of Eq.\ref{ham11}}
\label{fig1}
\end{figure}

As can be seen in Fig.\ref{fig1}, at high $T$ two characteristic length
scales occur. One is the charge ordering correlation length, $\xi $, and the
other, $l_{D}\sim \xi ^{-1}Q^{-1/2}$, is the effective Debye screening length%
$\ $of charged regions of size $\xi $ (see appendix C). Here we can already
recognize the effect of the competition between the short-range ordering
interaction and the long-range Coulomb interaction: the charge density is
homogeneous within regions of size $\xi $, but behaves like a plasma with
screening length $l_{D}\gg \xi $ on larger scales.\ In Ref.\cite{KKZNT95} it
was argued that the emergence of the screening length $l_{D}$ supports the
formation of compact ordered domains of size $l_{D}$ which then give rise to
a slow motion and thus glassiness. How such Debye screening should cause
such compact domains and glassiness has not been made explicit, however. Our
replica approach gives no indication for glassiness in the temperature
regime where the Debye screening theory applies. Thus, despite using the
same Hamiltonian, the stripe glass phase discussed in this paper is
qualitatively different from the scenario of Ref. \cite{KKZNT95}. A more
detailed discussion of these aspects is given in appendix C.

As the temperature is lowered, $\xi $\ increases monotonically while $l_{D}$
decreases until it becomes of the order of $\xi $. At this point, the Debye
screening approximation breaks down and the system crosses over to a new
regime characterized by spatial charge modulations, called stripes, with
period $l_{m}\simeq 2\pi Q^{-1/4}$ and coherence length $\xi $ \cite{NRK99}.
These modulations are particularly relevant at low temperatures where the
correlation length $\xi $ is larger than the inter-stripe distance $l_{m}$
and, as we will show, where the {\em stripe glass} phase\ emerges.

The charge correlation function ${\cal G}\left({\bf x,x}^{\prime }\right)
=T^{-1}\left\langle \varphi \left( {\bf x}\right) \varphi \left( {\bf x}%
^{\prime }\right) \right\rangle $ of the liquid state at low temperatures is
then characterized by two length scales, $l_{m}$ and $\xi $. After
transformation into momentum space, this function obeys the following
scaling behavior (neglecting effects due to anomalous powers): 
\begin{equation}
{\cal G}\left( {\bf q}\right) =l_{m}^{2}g(ql_{m},l_{m}/\xi ),
\label{GF_liquid}
\end{equation}
with ${\cal G}\left( {\bf q}\right) $ peaked at the modulation wave vector $%
q_{m}=\frac{2\pi }{l_{m}}$ with broadening $\ \xi ^{-1}$. \ In thermodynamic
equilibrium, the model, Eq.\ref{ham11} undergoes a stripe liquid - stripe
solid transition at some temperature $T_{c}$. Within a spherical
approximation\cite{CEK96} or a large-$N$ approximation \cite{NRK99}, the
transition is of second order and $T_{c}\rightarrow 0$ (unless one takes
additional lattice corrections into account which yield a finite $T_{c}$\cite
{NRK99}). In the Ising limit ($N=1$), $T_{c}>0$ and there are indications
that the transition is driven first order by fluctuations\cite{SAB75}.

Our results indicate that this phase transition may not be reached
kinematically in which case the system undergoes a glass transition instead.
In fact, according to our theory, glassiness emerges if the inter-stripe
correlations in the stripe liquid phase are sufficiently strong;
specifically if the ratio $\xi /l_{m}$ is larger than a critical value which
we find to be close to $2$. The temperature, $T_{A}$, where this happens,
within large-$N$ approximation of Eq.\ref{ham11}, is given by 
\begin{equation}
T_{A}=\frac{T_{c}^{0}}{\pi ^{2}Q^{1/4}+1},
\end{equation}
which decreases for increasing frustration parameter $Q$ .

Note that the criterion $\xi /l_{m}\simeq 2$ for glassiness is likely to be
much more general than the specific formula for $T_{A}$, which depends on
details of the model. Also, in a more realistic model, the stripe liquid -
stripe solid transition is expected to be of first order due to an
additional term $\sim \varphi ^{3}$ in Eq.\ref{ham11} which exists if
particle hole symmetry is broken. Nevertheless, our results do not depend on
an actual divergence of $\xi $ but solely that it is larger than a few
inter-stripe separations. Therefore, our theory also applies if the
transition is only moderately first order and the stripe solid does not
occurs unless $\xi >2l_{m}$.\ In this case, the equilibrium transition is
avoided and a glassy state results.

Our theory yields that below $T_{A\text{ }}$ the system establishes an
exponentially large number of metastable states and long time correlations,
characterized by the correlation function ${\cal F}\left( {\bf x,x}^{\prime
}\right) =T^{-1}\lim_{t\rightarrow \infty }\left\langle \phi \left( {\bf x}%
,t\right) \phi \left( {\bf x}^{\prime },0\right) \right\rangle $. These long
time correlations occur even though no state with actual long range order
exists. Even more interestingly, long time correlations with ${\cal F}\left( 
{\bf x,x}^{\prime }\right) \neq 0$ are unambiguously tied to a new length
scale, $\lambda $, which characterizes the typical length over which defects
and imperfections in the stripe pattern are allowed to wander over long
times. This length can be associated with the allowed vibrational motions in
a potential minima of the complex energy landscape of the system. In analogy
with structural glasses we therefore call it the Lindemann length of the
stripe glass.

Evidently, in the liquid state $\lambda $ is infinite. In the glassy state
we find that $\lambda $ jumps discontinuously to a finite value $\lambda
_{A}\simeq \xi \left( T_{A}\right) /3$ and continuously decreases at lower
temperature. The discontinuous jump in $\lambda $, unaccompanied by a latent
heat, indicates the transition is of the random first order type\cite{KTW89}%
, i.e., has one step replica symmetry breaking. Thus, the glassy state is
stable only if the slow motion of glassy textures is confined to a range
smaller than $\lambda _{A}$ (justifying our term Lindemann-length). Due to
this additional length scale the following scaling behavior of the long time
correlations results: 
\begin{equation}
{\cal F}\left( {\bf q}\right) =l_{m}^{2}j\left( ql_{m},\frac{l_{m}}{\xi },%
\frac{l_{m}}{\lambda }\right) .  \label{F_gen}
\end{equation}
with $j\left( x,y,z\right) =g\left( x,y\right) -g\left( x,\sqrt{z^{2}-y^{2}}%
\right) $ and $g$ from Eq.\ref{GF_liquid}.
\vskip 0.75cm  
\begin{figure}
\centerline{\epsfig{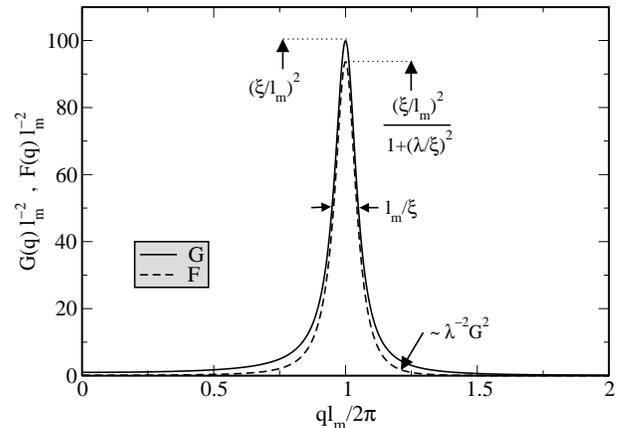}} 
\caption{The ${\cal G}$ and ${\cal F}$ \ correlation functions as given by 
Eqs. \ref{GF_liquid} and \ref{F_gen} for $\frac{l_{m}}{\xi}=\pi/10$ and
$\frac{l_{m}}{\lambda }=4\pi/5$.}
\label{fig2}
\end{figure}
In Fig.\ref{fig2} we show the momentum dependence of ${\cal F}\left( {\bf q}%
\right) $ in comparison with ${\cal G}\left( {\bf q}\right) $. Close to $%
q_{m}\equiv 2\pi /l_{m}$ (if $\left| \left| {\bf q}\right| -q_{m}\right|
\leq \xi ^{-1}$) we have ${\cal F}\left( {\bf q}\right) $ $\sim {\cal G}%
\left( {\bf q}\right) $. Configurations which are close to perfect stripe
arrangements are solely characterized by a momentum independent Debye-Waller
factor, such that ${\cal F}\left( {\bf q}\right) $ $\simeq \frac{1}{1+\left(
\lambda /\xi \right) ^{2}}{\cal G}\left( {\bf q}\right) $. On the other
hand, if $\left| \left| {\bf q}\right| -q_{m}\right| \gtrsim \lambda ^{-1}$
long time correlations are much reduced compared to instantaneous
correlations with ${\cal F}\left( {\bf q}\right) $ $\simeq \lambda ^{-2}%
{\cal G}\left( {\bf q}\right) ^{2}$. \ These `tails' of the correlation
functions are obviously built up by configurations with defects and
imperfections of the perfect stripe arrangement. Therefore, $\lambda $ has
to be interpreted as the length scale over which defects of the stripe
pattern are allowed to wander after a long time. The glassy state can only
be supported if $\lambda <\lambda _{A}$, it melts if defects are allowed to
wander too far.

Glassiness, including a viscous, energy-landscape dominated long time
relaxation, sets in due to the occurrence of exponentially many metastable
states ${\cal N}_{ms}\propto \exp \left( S_{c}\right) $.\cite{KTW89} We find
that the configurational entropy, 
\begin{equation}
S_{c}\left( T_{A}\right) \sim Q^{3/4}V\sim l_{m}^{-3}V,
\end{equation}
where $V$ is the volume of the system. The shorter the modulation length,
the larger is the number of possible metastable states, which is plausible
for simple geometrical reasons. This clearly demonstrates that locally the
stripe correlations stay intact in all these configurations. Furthermore, it
is the packing of stripes with different orientation and the arrangement of
defects that distinguishes the many different metastable states.

In the laboratory, the system will freeze into a glass, at some temperature $%
T_{G}<T_{A}$ which depends on the cooling rate. While this glass transition
is purely dynamical and distinct from a conventional phase transition, a key
feature of the ideal glass transition scenario of Ref. \cite{KTW89} is that
the slowing arises from proximity to an underlying random first order
transition at $T_{K}<T_{G}$, where the configurational entropy vanishes like 
$S_{c}(T)\propto T-T_{K}\ $. \ Our theory gives exactly this behavior with 
\[
S_{c}\left( T\right) =l_{m}^{-3}\Psi \left( \lambda /\xi ,l_{m}/\xi \right)
\,V. 
\]
We find $\Psi \left( s,t\right) =\left( 1+s^{-1}\right) \left( 1-t\right)
^{2}+\ln (1-\left( 1-t\right) ^{2})$ which vanishes linearly at a
temperature $T_{K}$. Below $T_{K}$ \ the system freezes into an amorphous
solid state due to this `entropy crisis'\cite{AWK48}, even for an infinitely
slow cooling rate. Usually, the reason for the system to prefer the liquid
state over the solid is entropic. If $S_{c}\to 0$ there is no entropic
advantage anymore to be in the liquid state and the amorphous solid results,
even in equilibrium.

Freezing into a glassy state below $T_{A}$ implies that within our replica
approach the barriers between different metastable states are infinite. This
however is a consequence of the mean field character of the replica
technique. Following Ref.\cite{KTW89} we argue that the formation of a
mosaic pattern with dynamically defined droplets of size $R$ of different
metastable states will occur. Entropy driven transitions between different
states lead to dynamical processes with relaxation time obeying a
Vogel-Fulcher law: 
\begin{equation}
\tau \varpropto \exp \left( \frac{DT_{K}}{T-T_{K}}\right) ,  \label{VFL}
\end{equation}
with fragility 
\begin{equation}
D\left( Q\right) \varpropto \frac{\sigma _{0}^{2}\left( Q\right) }{\left. 
\frac{dS_{c}\left( Q,T\right) }{dT}\right| _{T_{K}}}
\end{equation}
determined by the configurational entropy as well as the bare surface
tension of entropic droplets $\sigma _{0}\left( Q\right) $. Finally, a
simple estimate for $\sigma _{0}$, based on a variational argument, gives $%
D\left( Q\right) \varpropto \sqrt{Q}$ to a good approximation. This was
recently found in numerical simulations of the lattice version of Eq.\ref
{ham11} by Grousson {\em et al.} \cite{GT01}

\begin{figure}
\centerline{\epsfig{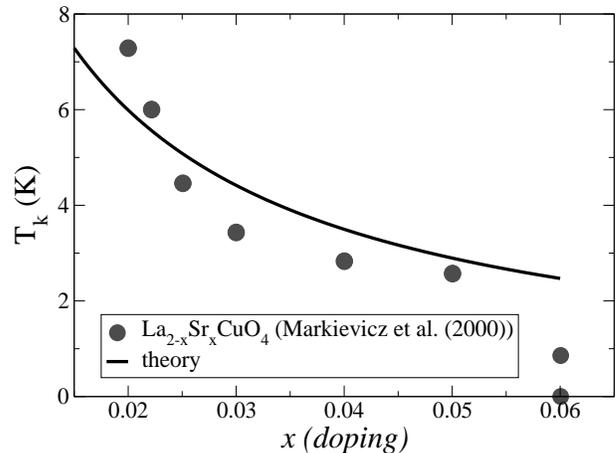}}
\caption{Comparison of $T_{K}$ with the experimental data analysed by
Markiewicz {\em et al.}$^{19}$}
\label{fig3}
\end{figure}

Recently, Markiewicz {\em et al.} \cite{MCPC00}{\em \ }analyzed various
experiments performed on $La_{2-x}Sr_{x}CuO_{4}$, spanning altogether 13
orders of magnitude of frequency, and also found a ''universal'' behavior of
the activation energies in underdoped cuprates. Interestingly, a good
description of the relaxational dynamics using a Vogel-Fulcher law, Eq.\ref
{VFL}, is possible. The analysis \ of Ref.\cite{MCPC00} also yields that $%
T_{K}$ decreases with increasing doping concentration $x$. Using the
relation, $l_{m}\simeq a\left( 2x\right) ^{-1}$,\cite{Y98} between
inter-stripe distance, $l_{m}$, and the doping concentration, as well as $%
l_{m}\simeq 2\pi Q^{-1/4}$ \ enables us to determine the doping dependence
of $T_{K}$ and show that it is properly described within our theory. This is
shown in Fig.\ref{fig3} in comparison with the results as deduced from
experiment in Ref. \cite{MCPC00}. Indeed, our theory gives the proper doping
dependence of $T_{K}$. If we neglect the differences between $T_{K}$ and $%
T_{A}$, (see inset of Fig.\ref{fig4} below), an approximate formula for the
doping dependence of $T_{K}$ is: $T_{K}\simeq \frac{T_{c}^{0}}{4\pi ^{3}x+1}$%
. The typical doping concentration on which changes in the glass transition
temperature occur is $x_{0}=\frac{1}{4\pi ^{3}}\simeq 0.008$. Note that the
experiments analyzed in Ref. \cite{MCPC00} are solely sensitive to the spin
excitations of the system. We argue that due to strong coupling between
charge and spin degrees of freedom (which is evident from the formation of
phase or anti-phase domain walls), glassiness of the charge density causes
the observed anomalous long time dynamics in the spin channel.
\vskip 0.75cm

\section{The Stripe Liquid}

In this section we summarize the main results for the stripe liquid state
needed \ for our subsequent calculations in the glass state. We will mostly
use results obtained within a the leading contribution of a $1/N$ expansion.
We calculate higher $1/N$ corrections and show that they are small in the
low temperature regime discussed here. Note also that the numerical solution
presented in Ref.\cite{SW00} did take higher $1/N$ corrections
systematically into account. The agreement between the results of this
chapter with these numerical results also supports the neglect of $1/N$
corrections for the stripe liquid state. The latter will become essential,
however, in the glass state.

The mean field equations of the stripe liquid including $1/N$ corrections
have been discussed in detail in Ref.\cite{NRK99}. Some of these results
have already been summarized above. In what follows we will mostly discuss
the low temperature regime where the correlation length, $\xi $, exceeds the
modulation length, $l_{m}$, of the system. This means we will be
concentrated in the high temperature region of the Fig.\ref{fig1}.

In the mean field approximation the correlation function is given by: 
\begin{equation}
{\cal G}\left( q\right) =\frac{1}{r+q^{2}+\frac{Q}{q^{2}}}  \label{GF_HF}
\end{equation}
where the parameter $r=r_{0}+u\left\langle \phi ^{2}\right\rangle $ must be
determined self consistently. At high temperatures, $r>2\sqrt{Q}$ \ and the
system is characterized by a correlation length $\xi \sim r^{-1/2}$, similar
to the unfrustrated system, as well as a Debye screening length $l_{D}\sim
\xi ^{-1}Q^{-1/2}$ characterizing conventional screening of charged objects
with linear size, $\xi $ and charge $\sim Q^{1/2}$. \ At lower temperatures,
when $r(T)<2\sqrt{Q}$, simple Debye screening breaks down and the system
establishes modulated structures with modulation length (inter stripe
distance), $l_{m}=4\pi /\sqrt{2\sqrt{Q}-r}$ and correlation length , $\xi =2/%
\sqrt{r+2\sqrt{Q}}$.\cite{NRK99}

However, unless $\xi $ becomes larger than $l_{m}$ no actual stripe
correlations emerge. This happens only if, at even lower temperatures, the
charge correlations are sufficiently strong to form a stripe liquid and
different stripes are strongly correlated. \ We will focus on this
temperature regime. Here, $0>r(T)>-2\sqrt{Q}$ and thus $G(x)$ exhibits an
oscillatory behavior with $\xi >l_{m}$.

It will be useful to introduce the positive dimensionless parameter $%
\varepsilon $ via 
\begin{equation}
\varepsilon ^{2}=\frac{4Q}{r^{2}}-1\text{ \ .}  \label{eps}
\end{equation}
We then find (defining $q_{m}^{2}=-r/2\sim Q^{1/2}$ with $l_{m}=\frac{2\pi }{%
q_{m}}$) that one can approximate the correlation function as 
\begin{equation}
{\cal G}({\bf q})\simeq \frac{q_{m}^{-2}\ }{\left( \left( \frac{q}{q_{m}}%
\right) ^{2}-1\right) ^{2}+\varepsilon ^{2}}\text{ .}  \label{G_app}
\end{equation}
In the last step we took only the leading term close to the peak at $q_{m}$
into account. By doing this approximation we are breaking the charge
neutrality condition (${\cal G}(q=0)=0$), only by a factor ${\cal G}(q=0)/%
{\cal G}(q_{m})=\varepsilon ^{2}$. For $\varepsilon \ll 1$ , \ the leading
contribution of the Fourier transform of ${\cal G}$ is given by 
\begin{equation}
G\left( x\right) =\frac{e^{-x/\xi \ }\sin (2\pi x/l_{m})}{4\pi \varepsilon
\left| x\right| }\text{,}
\end{equation}
which clearly shows the physical nature of the two length scales, $\xi =%
\frac{2}{\varepsilon q_{m}}$, and $l_{m}=\frac{2\pi }{q_{m}}$. In order to
have well correlated stripes, the correlation length $\xi \ $ must be much
bigger than the modulation length $l_{m}$. This \ indeed translates into $%
\varepsilon \ll 1$ which can be used as a small parameter of the theory.

Within the large-$N$ approximation, the temperature dependence of $r$ is
determined by 
\begin{equation}
r=r_{0}+u_{0}T\int \frac{d^{3}p}{8\pi ^{3}}{\cal G}\left( p\right) .
\label{mf}
\end{equation}
For the case without frustration, $Q=0$, the usual critical temperature $%
T_{c}^{0}=\frac{2\pi ^{2}\left| r_{0}\right| }{u\Lambda }$ results from the
requirement $r\left( T_{c}^{0}\right) =0$. However, for finite $Q$ no finite
transition temperature occurs within the large-$N$ approximation. Instead,
one finds from Eq.\ref{G_app} that 
\begin{equation}
r\left( T\right) =r_{0}+\frac{u_{0}T}{2\pi ^{2}}\left( \frac{\pi }{2}\frac{%
q_{m}}{\varepsilon }+\Lambda \right) .  \label{r_mf}
\end{equation}
In Eq.\ref{mf} we ignored additional $1/N$ corrections, characterized by the
self energy at zero momentum. In the limit $\xi >l_{m\text{ }}$ under
consideration we find $\Sigma _{G}\left( 0\right) \ \simeq -\frac{%
8q_{m}^{2}\varepsilon }{\pi }$. Since $\varepsilon $ is small, the one loop
self energy correction to $G$ can be safely neglected, given that $\Sigma
_{G}$ is an additive correction to the second term on the r.h.s. of Eq.\ref
{r_mf} which behaves, in leading order, as $\varepsilon ^{-1}$. \ From Eq.%
\ref{eps} and Eq.\ref{r_mf} we find at low temperatures 
\begin{equation}
\varepsilon =\frac{\pi }{2}Q^{1/4}\frac{T/T_{c}^{0}}{\frac{2\sqrt{Q}}{r_{0}}%
+1-\frac{T}{T_{c}^{0}}}\ \simeq \frac{\pi }{2}Q^{1/4}\frac{T/T_{c}^{0}}{1-%
\frac{T}{T_{c}^{0}}},  \label{eps_liq}
\end{equation}
where in the last step $\left| r_{0}\right| \gg 2\sqrt{Q}$ was assumed. This
relationship will be useful for the determination of the temperature, $T_{A}$%
, where glassiness sets in.

\section{The Stripe Glass}

\subsection{spontaneous ergodicity breaking and replica formalism}

With few exceptions\cite{CIS95,FH95,CKMP96,BCKM96}, the analytical
investigation of glassiness due to the emergence of a large number of
metastable states has concentrated on systems with quenched randomness. A
major step forward was made in Ref. \cite{Mon95,MP991}, where a new replica
approach, equally applicable to quenched random and nonrandom systems, was
developed. Within this approach, the configurational entropy for models of
structural glasses was \ calculated, in good agreement with computer
simulations.\cite{MP991} Here, we use this approach to investigate the
physics of self-generated stripe glasses.

The equilibrium's free energy density is given as $F=-\frac{T}{V}\log Z$. It
is of relevance only if the system is kinetically able to explore the entire
phase space. Alternatively, one can introduce the averaged typical free
energy of a system using the following recipe of an ``ergodicity breaking
field''\cite{Mon95,MP991} .

Locally stable field configurations can be identified using a test field $%
\psi \left( {\bf r}\right) $ and computing the partition sum 
\begin{equation}
Z\left[ \psi \right] =\int D\varphi e^{-{\cal H}[\varphi ]/T-\frac{g}{2}\int
d^{d}x\left[ \psi \left( {\bf x}\right) -\varphi \left( {\bf x}\right) %
\right] ^{2}},  \label{Zsig}
\end{equation}
where $g>0$ denotes the strength of the coupling. Evidently, the free energy 
\begin{equation}
\widetilde{f}\left[ \psi \right] =-T\log Z\left[ \psi \right]
\end{equation}
will be small when the field $\psi $ equals to a field configuration which
locally minimizes ${\cal H}$. Thus, sampling all configurations of the $\psi 
$-field, weighted with $\exp \left( -\beta \widetilde{f}\left[ \psi \right]
\right) $, is essentially a procedure to scan all locally stable
configurations. \ The quantity 
\begin{equation}
\widetilde{F}=\lim_{g\rightarrow 0}\frac{1}{W}\int D\psi \text{ }\widetilde{f%
}\left[ \psi \right] \exp \left( -\beta N\widetilde{f}\left[ \psi \right]
\right)  \label{f_tilda}
\end{equation}
is the weighted average of the free energy density of all locally stable
configurations. \ Here, $W=\int D\psi $ $\exp \left( -\widetilde{f}\left[
\psi \right] /T\right) $ is introduced for proper normalization.

It is physically appealing then to introduce the free energy difference, $%
\delta F$, via 
\begin{equation}
F=\widetilde{F}-\delta F,
\end{equation}
where $\delta F$ gives the amount of energy lost if the system is trapped
into locally stable states and hence not able to explore the entire phase
space of the ideal thermodynamic equilibrium. If the limit $g\rightarrow 0$
on Eq. \ref{f_tilda} behaves perturbatively, $\delta F=0$. This indicates
that the number of locally stable configurations stays finite in the
thermodynamic limit, or at least grows less rapid than exponential with $V$.
In this case all states are kinetically accessible. On the other hand, if
the limit $g\rightarrow 0$ does not behave perturbatively, it means that the
number of locally stable states,\ ${\cal N}_{{\rm ms}}$,$\ $\ is
exponentially large in $V$. This allows us to identify the difference
between the equilibrium and typical free energy as an entropy: 
\begin{equation}
\delta F=TS_{c}.  \label{class}
\end{equation}
The configurational entropy density, $S_{c}=\log {\cal N}_{{\rm ms}}$, is a
measure of the number of metastable states and is an extensive quantity if
there are exponentially many of those states. \ Its emergence renders the
system incapable of exploring the entire phase space. $S_{c}$ is then the
amount of entropy which the system that freezes it into a glassy state loses
due to its nonequilibrium- dynamics.

In order to find an explicit expression for $S_{c}$ one introduces a
replicated free energy: \cite{Mon95} 
\begin{equation}
F\left( m\right) =-\lim_{g\rightarrow 0}\frac{T}{m}\log \int D\ \psi Z\left[
\psi \right] ^{m}  \label{fofm}
\end{equation}
from which $\widetilde{F}$ can be obtained as $\widetilde{F}=\left. \frac{%
\partial mF(m)}{\partial m}\right| _{m=1}$and hence: 
\begin{equation}
S_{c}=\left. \frac{1}{T}\frac{\partial F(m)}{\partial m}\right| _{m=1}.
\label{conf1}
\end{equation}
Inserting $Z\left[ \psi \right] $ of Eq. \ref{Zsig} into Eq. \ref{fofm} and
integrating over $\psi ,$ one gets 
\begin{equation}
F\left( m\right) =-\frac{T}{m}\log Z(m)
\end{equation}
with replicated partition function given by 
\begin{eqnarray}
Z(m) &=&\lim_{g\rightarrow 0}\int D^{m}\varphi \exp \left( -\sum_{a=1}^{m}%
{\cal H}\left[ \varphi ^{a}\right] /T\right.  \nonumber \\
&&\text{ \ }\left. -\frac{g}{2m}\sum_{a,b=1}^{m}\int d^{d}x\varphi ^{a}({\bf %
x})\varphi ^{b}({\bf x})\right) ,  \label{repl_part}
\end{eqnarray}
which has a structure similar to a conventional equilibrium partition
function. The ergodicity breaking field $\psi $ causes a coupling between
replicas which might spontaneously lead to order in replica space even as $%
g\rightarrow 0$. This order is then associated with \ a finite $S_{c}$ and
thus glassiness.

Formally, Eq.\ref{repl_part} equals the partition function of system with
quenched random field analyzed using the conventional replica approach. The
main difference is that, here, the limit $m\rightarrow 1$ has to be taken.
The resulting many body problem in replica space is characterized by the
matrix correlation function, ${\cal G}_{ab}\left( {\bf q}\right)
=\left\langle \varphi ^{a}({\bf q)}\varphi ^{b}(-{\bf q)}\right\rangle $, in
replica space with Dyson equation: 
\begin{equation}
\left. {\cal G}^{-1}\left( {\bf q}\right) \right| _{ab}={\cal G}%
_{0}^{-1}\left( {\bf q}\right) \delta _{ab}+\Sigma _{ab}\left( {\bf q}%
\right) -\frac{g}{m}.  \label{dyson}
\end{equation}
Here, ${\cal G}_{0}\left( {\bf q}\right) $ is the Hartree propagator of Eq.%
\ref{GF_HF} which we approximate at low temperatures by Eq.\ref{G_app}. $%
\Sigma _{ab}\left( {\bf q}\right) $ is the self energy in replica space. If
we find that as a consequence of the ergodicity breaking coupling constant, $%
g$, $\Sigma _{ab}\left( {\bf q}\right) $ has finite off diagonal elements we
can conclude that there must be an energy landscape sensitive to the
infinitesimal perturbation, $g$, supporting a glassy dynamics. On the other
hand, if $\Sigma _{ab}\left( {\bf q}\right) $ is diagonal, conventional
ergodic dynamics results and the system is in its liquid state or may build
an ordered solid. As pointed out above, this strategy is similar to the
investigation of symmetry breaking in conventional phase transitions where
the off diagonal elements of an appropriately defined matrix self energy are
associated with the order parameter of the transition (superconducting gap
function or staggered magnetization in case of a superconductor or
antiferromagnet, respectively). However, it turns out that in the present
case the off diagonal elements of $\Sigma _{ab}\left( {\bf q}\right) $ jump
discontinuously from zero to a finite value and a linearized theory with $%
\Sigma _{ab}\left( {\bf q}\right) \left( 1-\delta _{ab}\right) $ small
(which determines the transition temperature in case of continuous phase
transitions) will only give the trivial solution with vanishing off diagonal
elements. A nonlinear theory for $\Sigma _{ab}\left( {\bf q}\right) $ needs
to be developed.

Since the attractive potential between different replicas is symmetric with
respect to the replica index we use the following Ansatz for the Green's
function 
\begin{equation}
{\cal G}_{ab}\left( {\bf q}\right) =\left( {\cal G}\left( {\bf q}\right) -%
{\cal F}\left( {\bf q}\right) \right) \delta _{ab}+{\cal F}\left( {\bf q}%
\right) ,  \label{repans}
\end{equation}
i.e. with equal diagonal elements, ${\cal G}\left( {\bf q}\right) $, and
equal off-diagonal elements, ${\cal F}\left( {\bf q}\right) $. Note, if one
applies the present replica formalism to systems with quenched disorder it
turns out that the replica symmetric Ansatz, Eq.\ref{repans}, is equivalent
to one-step replica symmetry breaking in the conventional replica formalism 
\cite{Mon95}. The physical interpretation of ${\cal G}\left( {\bf r-r}%
^{\prime }\right) =T^{-1}\left\langle \varphi ({\bf r)}\varphi ({\bf r}%
^{\prime }{\bf )}\right\rangle $ as thermodynamic (instantaneous)
correlation function is straightforward. On the other hand, ${\cal F}\left( 
{\bf r-r}^{\prime }\right) =T^{-1}\lim_{t\rightarrow \infty }\left\langle
\varphi ({\bf r,}t{\bf )}\varphi ({\bf r}^{\prime },0{\bf )}\right\rangle $
can be interpreted as measuring long time correlations. As shown in the
appendix A, \ inserting the Ansatz, Eq.\ref{repans}, into Eq.\ref{dyson}
gives in the limit $m\rightarrow 1$: 
\begin{equation}
{\cal G}^{-1}\left( {\bf q}\right) ={\cal G}_{0}^{-1}\left( {\bf q}\right)
+\Sigma _{{\cal G}}\left( {\bf q}\right)  \label{Gdyson}
\end{equation}
for the diagonal elements, and 
\begin{eqnarray}
{\cal F}\left( {\bf q}\right) &=&\frac{-{\cal G}^{2}\left( {\bf q}\right)
\Sigma _{{\cal F}}\left( {\bf q}\right) }{1-{\cal G}\left( {\bf q}\right)
\Sigma _{{\cal F}}\left( {\bf q}\right) }  \nonumber \\
&=&{\cal G}\left( {\bf q}\right) -\frac{1}{{\cal G}^{-1}\left( {\bf q}%
\right) -\Sigma _{{\cal F}}\left( {\bf q}\right) }  \nonumber \\
&\equiv &{\cal G}\left( {\bf q}\right) -{\cal K}\left( {\bf q}\right)
\label{Dys_od}
\end{eqnarray}
for the off diagonal elements, respectively. Here, $\Sigma _{{\cal G}}$ and $%
\Sigma _{{\cal F}}$ are the diagonal and off diagonal elements of the self
energy in replica space. \ In the last equation ${\cal K}\ $denotes the
deviations of the long time and instantaneous correlations. Analyzing the
corresponding dynamical equations of the problem, it turns out that ${\cal K}
$ is the static retarded response function\cite{WSW01b}. In the liquid state
fluctuation dissipation theorem gives ${\cal K}={\cal G}$ and no long time
correlations occur.

\subsection{Defect wandering in stripe glasses}

The self energy in replica space was numerically investigated in Ref. \cite
{SW00} within the self consistent screening approximation which we summarize
in appendix A. It was shown that below a characteristic temperature, $T_{A}$%
, an off diagonal self energy in replica space emerges, leading to finite
long time correlations, ${\cal F}\left( {\bf q}\right) $, as well as a
finite configurational entropy density 
\begin{equation}
s_{c}=S_{c}/V
\end{equation}
in the thermodynamic limit, $V\rightarrow \infty $. Here we present an
approximate but analytical solution of the same set of equations which has
the appeal that the underlying physics of the stripe glass formation becomes
much more transparent. It also reveals more directly the emergence of a new
length scale which characterizes the wandering of defects in the stripe
pattern after long times. In this context, we demonstrate that the melting
of the glass as $T$ becomes larger than $T_{A}$ is a consequence of the fact
that the characteristic length for defect wandering becomes too large; the
glass becomes too fluid causing the devitrification into a stripe liquid.

The key assumption of the analytical approach to the self consistent
screening approximation is that the off diagonal self energy $\Sigma _{{\cal %
F}}\left( {\bf q}\right) $ is weakly momentum dependent. Specifically, we
assume that due to the strong momentum dependence of the correlation
function, the product ${\cal G}\left( {\bf q}\right) \Sigma _{{\cal F}%
}\left( {\bf q}\right) $ varies with ${\bf q}$ predominantly due to ${\cal G}%
\left( {\bf q}\right) $. This assumption will be justified a posteriori.
Also, our numerical results, which were obtained without any restriction on
the ${\bf q}$-dependence of $\Sigma _{{\cal F}}\left( {\bf q}\right) $,
clearly show that this assumption is justified. We will then calculate $%
\Sigma _{{\cal F}}(q_{m})$ at the modulation wave vector $q_{m}$.

\begin{figure}
\centerline{\epsfig{file=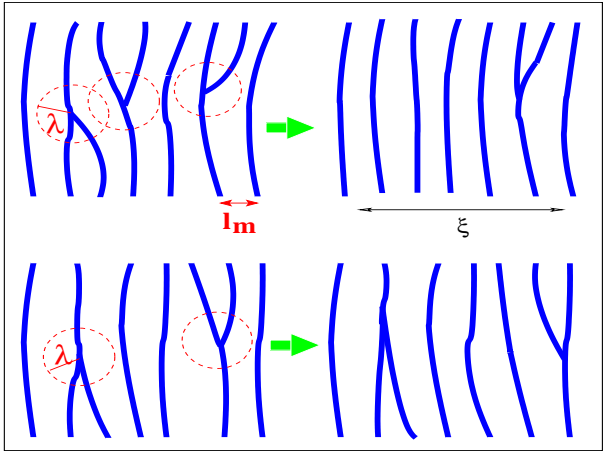,width=8cm,height=6cm,scale=0.85}}
\caption{Pictorial description of the wandering of defects in the stripe pattern. The 
upper panel shows defects that can be healed by the wandering process. The lower pannel 
shows defects which are too far apart and cannot be healed.}
\label{deffectwander}
\end{figure}

\mbox{$>$}%
>From the analysis of the liquid state we know that ${\cal G}(q)$ is strongly
peaked at the modulation vector $q_{m}$ with width $\xi ^{-1}$. By
inspection of the Dyson equation, Eq.\ref{Dys_od}, for ${\cal F}$, one can
see that for $q\sim q_{m}$ and $\Sigma _{{\cal F}}(q_{m}){\cal G}(q_{m})\gg
1 $ (since ${\cal G}$ is peaked around $q=q_{m}$): 
\begin{equation}
{\cal F}(q)\lesssim {\cal G}(q).  \label{FG1}
\end{equation}
Moreover, ${\cal G}$ vanishes rapidly away from the peak (as does $\Sigma
_{F}(q_{m})G(q_{m})$) and it follows from the same equation, \ref{Dys_od},
that for large $\left| q-q_{m}\right| \ $ 
\begin{equation}
{\cal F}(q)\simeq -\Sigma _{{\cal F}}(q){\cal G}^{2}(q).  \label{FG2}
\end{equation}
If a solution for ${\cal F}$ exists, it is going to be a peaked at $q_{m}$,
but smaller and narrower than ${\cal G}$. Consequently, if a stripe glass
occurs, the long time limit of the correlation function in not just a
slightly rescaled version of the instantaneous correlation function, but it
is multiplied by a $q$ dependent function that leads to a qualitatively
different behavior for different momenta. Once a glassy state is formed,
configurations which contribute to the peaks of ${\cal G}(q)$ and ${\cal F}%
(q)$, i.e. almost perfect stripe configurations, are almost unchanged even
after long times. Close to $q_{m}$ ${\cal F}(q)\,$\ is solely reduced by
some momentum independent Debye-Waller factor $\exp \left( -D\right) ={\cal F%
}(q)/{\cal G}(q)$. On the other hand, certain configurations which form the
tails of ${\cal G}(q)$, i.e. defects and imperfections of the stripe
pattern, disappear after a long time since now ${\cal F}(q)\ll {\cal G}(q)$.
The ratio of both functions is now strongly momentum dependent. ${\cal F}(q)$
becomes sharper than ${\cal G}(q)$ because certain defects got healed in
time. Evidently, there must be a momentum scale (or equivalently a length
scale) which determines the transition between these two regimes. In what
follows we will identify and determine this length scale. A pictorial
description of the defects on the stripe configuration and the meaning of $%
\lambda $ is presented on Fig.\ref{deffectwander}.

Due to our assumption that $\Sigma _{{\cal F}}$ is weakly dependent on $q$,
we concentrate on $\Sigma _{{\cal F}}\left( q_{m}\right) $ at the modulation
wave vector. One easily finds that $\Sigma _{{\cal F}}\left( q_{m}\right)
\leq 0$. A dimensional analysis furthermore shows that $\Sigma _{{\cal F}}$
is length$^{-2}$. This suggests to define a new length scale, $\lambda $,
via 
\begin{equation}
\ \Sigma _{{\cal F}}(q_{m})=-\ \left( \frac{2}{\lambda }\right) ^{2}.
\label{Sig_fans}
\end{equation}
For the subsequent calculation it is convenient to introduce in addition to
the dimensionless parameter $\varepsilon $ which gives $\xi ^{-1}=\frac{%
\varepsilon q_{m}}{2}$ a new dimensionless parameter, $\kappa $, defined via 
\begin{equation}
\lambda ^{-1}=\frac{\sqrt{\kappa ^{2}-\varepsilon ^{2}}q_{m}}{2}.
\label{lambda}
\end{equation}
Obviously, in the liquid state, where $\ \Sigma _{{\cal F}}\rightarrow 0$,
we find $\lambda \rightarrow \infty $ and it holds $\kappa =\varepsilon $.
In a glassy state $\kappa >\varepsilon $. This Ansatz for $\Sigma _{{\cal F}%
} $, inserted into Eq.\ref{Dys_od}, yields 
\begin{equation}
{\cal K}\left( q\right) =\frac{q_{m}^{-2}}{\left( \left( \frac{q}{q_{m}}%
\right) ^{2}-1\right) ^{2}+\kappa ^{2}}  \label{K_app}
\end{equation}
for the correlation function ${\cal K}={\cal G}-{\cal F}$. \ Note that $%
{\cal K}$ has the same structure as ${\cal G}$ but with $\varepsilon
\rightarrow \kappa $. It immediately follows that it is the length scale $%
\lambda $ which determines whether long time correlations are similar or
different from instantaneous ones. If $\left| q-q_{m}\right| <\lambda ^{-1}$
Eq.\ref{FG1} holds, whereas for $\left| q-q_{m}\right| >\lambda ^{-1}$ long
time correlations are strongly reduced leading to Eq.\ref{FG2}. Consequently
we identify $\lambda $ as the length scale over which imperfections of the
stripe pattern manage to wander, i.e. defects can be healed, even in the
frozen glass state.

\begin{figure}
\centerline{\epsfig{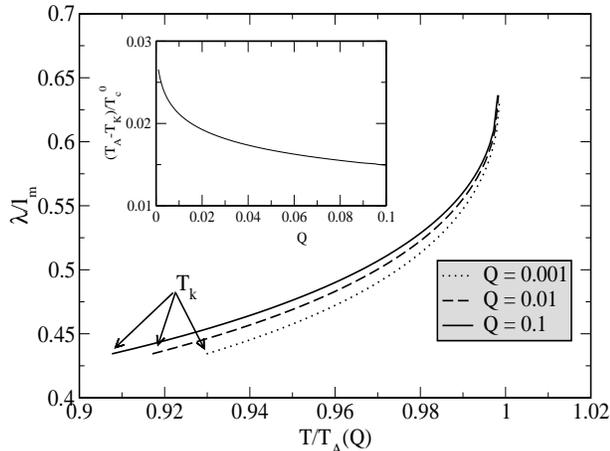}} 
\caption{The Lindeman length $\protect\lambda $ as a function of temperature
for different values of $Q$. Inset: $Q$ dependence of $T_A-T_K$}
\label{fig4}
\end{figure}

The next step is to determine $\Sigma _{{\cal F}}(q_{m})$ for a given value
of $\lambda $ and to self consistently determine this length scale. The
details of the calculation of $\Sigma _{{\cal F}}(q_{m})$ with ${\cal G}%
\left( q\right) $ as given in Eq.\ref{G_app} and ${\cal K}\left( q\right) $
of Eq.\ref{K_app} are summarized in the appendix B. The result is 
\begin{equation}
\Sigma _{{\cal F}}\left( q_{m}\right) =-\frac{8q_{m}^{2}\varepsilon ^{2}}{%
\pi }\frac{\left( 1-\frac{\varepsilon }{\kappa }\right) ^{2}}{1-\left( 1-%
\frac{\varepsilon }{\kappa }\right) ^{2}}\left( \frac{1}{\varepsilon }-\frac{%
1}{\kappa }\right) .  \label{sigf_calc}
\end{equation}
This has to be compared with our ansatz, Eq.\ref{Sig_fans}. Together with Eq.%
\ref{lambda} this gives 
\begin{equation}
\Sigma _{{\cal F}}\left( q_{m}\right) =-\left( \kappa ^{2}-\varepsilon
^{2}\right) q_{m}^{2}.  \label{h_lbl}
\end{equation}
Comparing Eqs.\ref{sigf_calc} and \ref{h_lbl} we immediately find the
nonlinear algebraic equation 
\begin{equation}
\kappa ^{2}-\varepsilon ^{2}=\frac{8\varepsilon ^{2}}{\pi }\frac{\left( 1-%
\frac{\varepsilon }{\kappa }\right) ^{2}}{1-\left( 1-\frac{\varepsilon }{%
\kappa }\right) ^{2}}\left( \frac{1}{\varepsilon }-\frac{1}{\kappa }\right)
\   \label{h_lbl2}
\end{equation}
which determines the length scale $\lambda $ (via $\kappa $) , as function
of the correlation length and the modulation length, i.e. properties of the
liquid state. One solution of this equation is always $\varepsilon =\kappa $%
, which corresponds to the liquid state. Factorizing this trivial solution
from \ref{h_lbl2} we find that the other solutions are given by 
\begin{equation}
\ \frac{\varepsilon }{\kappa }\left[ \frac{8}{\pi \varepsilon }\left( 1-%
\frac{\varepsilon }{\kappa }\right) ^{2}-\left( 1-\frac{\varepsilon }{\kappa 
}\right) \right] =2.  \label{criterion}
\end{equation}
This is a cubic equation which can be solved exactly. Before we discuss this
equation in some detail we analyze the condition for obtaining a nontrivial
solution which corresponds to the onset of glassiness. For $\frac{8}{\pi
\varepsilon }\gg 1$ the left hand side of Eq.\ref{criterion} has its maximum 
$\ \frac{1}{\pi \varepsilon }$ at $\kappa \simeq 3\varepsilon $ which gives
the condition for the existence of a solution as $\frac{1}{\pi \varepsilon }%
\geq 2.$ This defines the critical value of $\varepsilon $ as 
\begin{equation}
\varepsilon _{A}=\varepsilon \left( T_{A}\right) \simeq \frac{1}{2\pi }.
\end{equation}
Thus, if the ratio of the correlation length and the modulation length
exceeds the critical value 
\begin{equation}
\left. \frac{\xi }{l_{m}}\right| _{T=T_{A}}\simeq 2
\end{equation}
long time glassy correlations emerge. As expected, besides strong
frustrations, glassiness also requires sufficiently strong liquid
correlations. Since it follows from Eq.\ref{criterion} that $\kappa \simeq
3\varepsilon $ it also follows that 
\begin{equation}
\left. \frac{\lambda }{l_{m}}\right| _{T=T_{A}}\simeq \frac{2}{3}.
\end{equation}
This type of behavior is evident on Fig.\ref{fig4} where, independent of the
value of $Q$ all curves for $\lambda /l_{m}$ reach the same maximum value at 
$T_{A}$. If defects and imperfections of the stripe pattern within the
glassy state manage to flow over length scales larger than $\simeq \frac{2}{3%
}l_{m}$ the glass becomes unstable because it is too fluid to support a
frozen state. Thus, we identify the length scale $\lambda $ as the Lindemann
length of the glass.

As can be seen on Fig.\ref{fig4}, $\lambda $ is a monotonically increasing
function of temperature. At small temperatures $\lambda $ grows linearly,
evolving to a cusp at the dynamical freezing temperature, $T_{A}$. Above $%
T_{A}$, $\lambda $ becomes infinite and the devitrification is complete.

Using the $Q$-dependence of $\varepsilon $ in the liquid state (see Eq.\ref
{eps_liq}) the stripe glass - stripe liquid transition temperature is then
given by 
\begin{equation}
\varepsilon _{A}=\frac{1}{2\pi }=\frac{\pi }{2}Q^{1/4}\frac{T_{A}/T_{c}^{0}}{%
1-\frac{T_{A}}{T_{c}^{0}}}
\end{equation}
which gives 
\begin{equation}
T_{A}=\frac{T_{c}^{0}}{\pi ^{2}Q^{1/4}+1}
\end{equation}
where $T_{c}^{0}$ is critical temperature of the $Q=0$ problem. Moreover,
since the difference between $T_{A}$ and $T_{K}$ is small compared to $%
T_{c}^{0}$ (see inset of Fig.\ref{fig4}), the $Q$ dependence of $T_{K}$ will
be roughly the same as of $T_{A}$. As pointed out above, we do not expect $%
T_{A}$ to be sensitively affected by an additional $\frac{v}{3}\varphi ^{3}$%
-term in the Hamiltonian. This is less clear for the Kautzmann temperature
and the difference between $T_{A}$ and $T_{K}$ might well depend on the
coupling constant $v$. Still we expect $T_{K}$ to decrease for increasing $Q$%
.
\vskip 0.75cm

\subsection{On lattice corrections}

\begin{figure}
\centerline{\epsfig{file=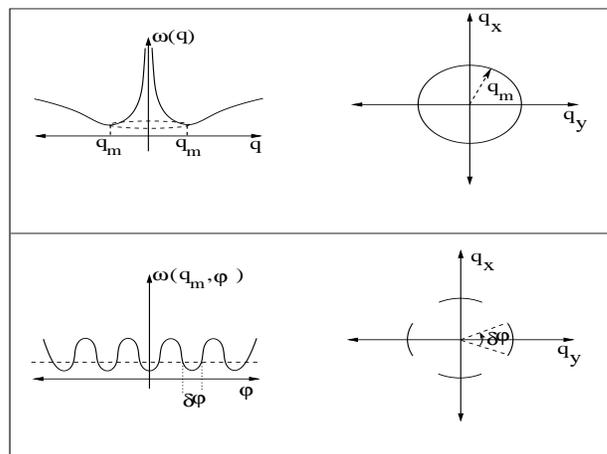,width=8cm,height=6cm,scale=0.85}} 
\caption{Low energy modes. Upper panel: in the continuum's limit, $\Delta=0$,
the phase space for low energy excitations is a d-1 sphere. Lower panel: The anisotropy 
gives rise to a ``direction dependent mass'' and the phase space for low energy excitations is reduced to a set of arcs}
\label{latt_corr}
\end{figure}

Until this point our theory has been performed in the continuum's limit and
effects due to lattice corrections have been neglected. The particular form
of the propagator, Eqs.\ref{GF_HF} and \ref{G_app}, \ however gives rise to
a specific sensibility of our results to lattice corrections which is worth
mentioning. Within the continuum's limit, the large-$N$ approach used in
this paper yields no ordinary phase transition to a stripe solid state, a
result which has been pointed out earlier \cite{NRK99,CEK96}. This can be
most easily seen from the mean field equation, Eq.\ref{r_mf}. A solution $%
\varepsilon =0$ is only allowed if $T=0$. For $T>0$ one always finds a
large, but finite correlation length, $\xi =\frac{2}{\varepsilon q_{m}}$.
The absence of an ordered state is a consequence of the dispersion relation $%
\omega _{q}$, with 
\begin{equation}
\omega _{q}=q_{m}\sqrt{\left( 1-\left( \frac{q}{q_{m}}\right) ^{2}\right)
^{2}+\varepsilon ^{2}}.  \label{mode}
\end{equation}
In contrast to an antiferromagnet or a conventional charge density wave,
where the low energy modes are determined by isolated points in momentum
space, Eq.\ref{mode} gives rise to a $d-1$ dimensional sphere of low energy
modes in momentum space (see upper panel of Fig.\ref{latt_corr}). It is this
large phase space of low energy excitations which destroys long range order.

The most dramatic effect of corrections beyond the continuum's limit is the
appearance of an anisotropy on the low energy states. It is reasonable to
assume that the Hamiltonian \ref{ham11} has a next order correction of the
type $4Kq_{x}^{2}q_{y}^{2}$ which leads to a direction dependent ``mass''
term in Eq.\ref{mode} 
\begin{equation}
\varepsilon ^{2}=\varepsilon ^{2}\left( K=0\right) +Kq_{m}^{2}\sin
^{2}\left( 2\phi \right)
\end{equation}
with angle $\phi $ (in the $x-y$ plane) and a dimensionless anisotropy
parameter $K$. Of course this is only justified if $K\ll q_{m}^{-2}$. If
however $K\simeq q_{m}^{-2}$ the physics strongly depends on phenomena on
the scale of the interatomic spacing.

The mean field equation for finite $K$ is then given by 
\begin{equation}
r\left( T\right) =r_{0}+\frac{u_{0}T}{2\pi ^{2}}\left( \frac{\pi }{2}\frac{%
q_{m}}{\sqrt{\varepsilon \left( K=0\right) ^{2}+Kq_{m}^{2}}}+1\right) .
\end{equation}
The system now undergoes a phase transition for arbitrary small $K$ at $%
T_{c}=\frac{T_{c}^{0}}{\pi K^{-1/2}\Lambda +1}$ . As expected, $T_{c}$
vanishes as $K\rightarrow 0$ . Moreover, as pointed out on Refs.\cite
{NRK99,CEK96}, $T_{c}$ does not merge with $T_{c}^{0}$ as $Q\rightarrow 0$.
The origin of this phase transition is that the $d-1$ dimensional sphere of
low energy degrees of freedom is reduced to arcs of size $\delta \varphi
\sim \varepsilon /(K^{1/2}q_{m})$ (see lower panel of Fig.\ref{latt_corr}).
As $\varepsilon $ decreases, these arcs become indistinguishable from
isolated points, a behavior similar to\ the case of an antiferromagnet or a
charge density wave occurs, leading to an ordinary phase transition. The
mean field analysis of the lattice version of Eq.?? of Ref. \cite{CEK96},
which finds $T_{c}$ considerably smaller than $T_{c}^{0}$ supports $K\ll
q_{m}^{-2}$.

In analogy to the mean field analysis of the liquid state one can also
perform the theory of the glassy state for finite $K$. If lattice
corrections are strong and $K\sim q_{m}^{-2}$ a transition to a stripe solid
occurs at $T_{c}$. The low energy excitations are located at isolated points
and the behavior is equivalent to the one of an unfrustrated system. In this
case, the glassy state will only occur if the solidification is avoided by
supercooling. On the other hand, if $K\ll q_{m}^{-2}$ the low energy modes
are unchanged and the lattice corrections become irrelevant. Since the glass
transition does not require $\varepsilon $ to vanish, but solely to reach a
certain finite limit $\sim \frac{1}{2\pi },$ we conclude that for $K\lesssim 
\frac{q_{m}^{-2}}{2\pi }$ the glass transition is essentially unchanged.

\subsection{The configurational entropy}

\begin{figure} 
\centerline{\epsfig{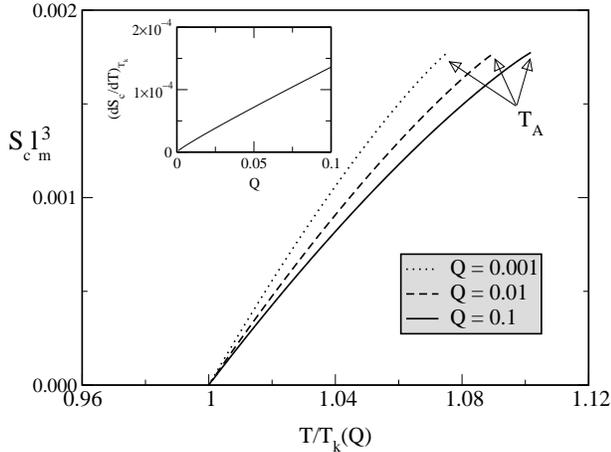}} 
\caption{The configurational entropy for different values of $Q$. Inset:
dependence of $\left( dS_{c}/dT\right) _{T_{k}}$ on $Q$}
\label{fig5}
\end{figure}

The main argument for the emergence of a glassy state is the occurrence of
an exponentially large number if metastable states, characterized by the
configurational entropy. $S_{c}$ is determined from Eq.\ref{conf1} from the
replicated theory defined in Eq.\ref{repl_part}, which gives $F\left(
m\right) =-\frac{T}{m}\log Z(m)$. Within the self consistent screening
approximation it follows 
\[
F\left( m\right) /\left( 2mT\right) ={\rm tr}\log {\cal G}^{-1}+{\rm tr}\log 
{\cal D}^{-1}-{\rm tr}\Sigma {\cal G} 
\]
Since all quantities are matrices in replica space with a structure given in
Eq.\ref{repans} the evaluation of expressions like ${\rm tr}\log {\cal G}%
^{-1}$ etc. becomes straightforward. Performing the derivative with respect
to the number of replicas according to Eq.\ref{conf1} gives immediately for
the entropy density:

\[
s_{c}=s_{c}^{(1)}+s_{c}^{(2)} 
\]
with the two contributions: 
\[
s_{c}^{(1)}=-\frac{1}{2}\int \frac{d^{3}q}{8\pi ^{3}}\left\{ \ln \left( 1-%
\frac{{\cal F}({\bf q})}{{\cal G}({\bf q})}\right) +\frac{{\cal F}({\bf q})}{%
{\cal G}({\bf q})}\right\} 
\]
and 
\[
s_{c}^{(2)}=\frac{1}{2}\int \frac{d^{3}q}{8\pi ^{3}}\left\{ \ln \left( 1-%
\frac{v_{0}\Pi _{{\cal F}}({\bf q})}{1+v_{0}\Pi _{{\cal G}}({\bf q})}\right)
+\frac{v_{0}\Pi _{{\cal F}}({\bf q})}{1+v_{0}\Pi _{{\cal G}}({\bf q})}%
\right\} . 
\]
For the definition of $\Pi _{{\cal F}}$, $\Pi _{{\cal G}}$ etc. see appendix
A. Using the same approximations as for the evaluation of the self energy, $%
\Sigma _{{\cal F}}$, in appendix B we find 
\[
\frac{{\cal F}({\bf q})}{{\cal G}({\bf q})}\ \approx \frac{\kappa
^{2}-\varepsilon ^{2}}{\left[ \left( \frac{q}{q_{m}}\right) ^{2}-1\right]
^{2}+\kappa ^{2}}. 
\]
The evaluation of the integrals is straightforward and we find 
\begin{eqnarray}
s_{c}^{(1)} &=&\frac{q_{m}^{3}}{4\pi }\frac{\kappa }{2}\left( 1-\frac{%
\varepsilon }{\kappa }\right) ^{2}  \nonumber \\
s_{c}^{(2)} &=&\frac{q_{m}^{3}}{4\pi }\frac{2}{\pi }\left\{ \left( 1-\frac{%
\varepsilon }{\kappa }\right) ^{2}+\ln (1-\left( 1-\frac{\varepsilon }{%
\kappa }\right) ^{2})\right\} .
\end{eqnarray}
Obviously, $s_{c}\neq 0$ only if $\kappa >\varepsilon $ i.e. the Lindemann
length, $\lambda $, is finite. $s_{c}$ vanishes in the liquid state where $%
\kappa =\varepsilon $. The results for $s_{c}\left( T\right) $ for different
values of $Q$ are given on Fig.\ref{fig5}

Using the results of the previous section where we found that for $T=T_{A}$
the dimensionless quantities $\varepsilon $ and $\kappa $ take fixed values,
\ we obtain at $T=T_{A}$: 
\begin{equation}
S_{c}\left( T_{A}\right) =\ CVQ^{3/4}  \label{scvsq}
\end{equation}
with $C=\left( \frac{2}{\pi }+\frac{3}{4\pi }\right) \frac{1}{9\pi }+\frac{1%
}{2\pi ^{2}}\ln (\frac{5}{9})=\allowbreak 1.\,\allowbreak 181\,6\times
10^{-3}\ $. \ The configurational entropy decreases for decreasing $Q$.
Since the modulation length behaves as $l_{m}\sim Q^{1/3}$ it follows $%
S_{c}\varpropto l_{m}^{-3}$. The larger the modulation length, the smaller
is the number of states one can form which clearly demonstrates that locally
stripe correlations stay intact in all these configurations. It is more the
packing and arrangement of defects which distinguishes different metastable
states. Less of those packings are possible per unit volume if the
modulation length grows, for simple geometrical reasons. The relation, Eq.%
\ref{scvsq} was recently derived by us using the concept of replica bound
states.\cite{SWW00} The fact that we obtain the same result using an
entirely different approach to solve the problem increases our confidence in
the applicability of the self consistent screening approximation, which also
allows us to calculate the constant $C$. We can also compare $S_{c}\left(
T_{A}\right) $ with the numerical results of Ref.\cite{SW00}. $S_{c}$ was
calculated in Ref. \cite{SW00} for the values $Q=0.01$ and $Q=0.001$, where
we found $S_{c}=3.45\times 10^{-5}$ and $6.4\times 10^{-6}$, respectively.
>From Eq.\ref{scvsq} we find the values $\ \allowbreak 3.\,\allowbreak
7\times 10^{-5}$ and $6.\,\allowbreak 6\times 10^{-6}$, which agrees well
with the numerical results.

Finally, the condition $S_{c}\left( T=T_{K}\right) =0$ determines the
Kauzmann temperature below which no entropic advantage of the liquid state,
compared to the amorphous solid state (the glass) exists. Even if one
manages to anneal the liquid down to $T_{K}$ without freezing into a glass,
something which might me achieved using an infinitely slow cooling rate, a
mandatory transition into an amorphous solid occurs at $T_{K}$. In Fig.\ref
{fig5} we show the temperature dependence of $S_{c}$ for different values of
the frustration parameter $Q$ as well as the $Q$ dependence of the slope $%
(dS_c/dT)_{T_K}$.

\subsection{Dynamics and flow via entropic droplet formation}

So far, we have shown that within the self consistent screening
approximation of Eq.\ref{ham11}, an exponentially large number of metastable
states occurs. The conclusion that somewhere below $T_{A}$ nonequilibrium
dynamic sets in is actually obtained using a purely thermodynamic
characterization of \ the spectrum of metastable states. It is very
plausible \ that an exponentially large number of metastable states is
necessarily connected to glassy dynamics. In several random spin models as
well as in the model of self generated glassiness in frustrated Josephson
junction arrays, this point of view has been clearly supported by actual
dynamical calculations. For so called infinite range models (i.e. within
mean field approximation), where the barriers between the various metastable
states diverges, freezing at $T_{A}$ has been found.\cite{KTW89,CS95}

Solving the Langevin equation for the time evolution of the correlation and
response function of $\varphi \left( {\bf x},t\right) $, using the
supersymmetric formulation of the Martin-Siggia-Rose approach\cite{ZJbook}
within the self consistent screening approximation, we find that the
emergence of exponentially many metastable states also leads to stripe glass
state within a dynamical approach.\cite{WSW01b} Interestingly, this mode
coupling type approach gives exactly the same criterion for the emergence of
glassiness as the above replica approach only if one properly takes the
aging behavior of the dynamical evolution into account, following Ref.\cite
{CK95}. Thus, the replica approach employed here takes the effects of aging
correctly into account.

Within the mode coupling or replica integral equation approaches a perfect
freezing occurs at the temperature $T_{A}$. However, more realistically, $%
T_{A}$ is rather a crossover scale to a regime with slow activated dynamics
and not the actual freezing temperature. Depending on the history of the
system, the laboratory glass transition where the time scale of motions
exceeds a certain limit occurs somewhere between $T_{K}$ and $T_{A}$. \ A
key question in this context is the nature of the dynamical processes for $%
T_{K}<T<T_{A}$.

This dynamics involves droplets or instantons, essential singularities from
the point of view of perturbative approaches. A complete formal treatment is
therefore difficult. Nevertheless, a reasonable description of the dynamics
in a system with finite $S_{c}$ is given in Ref.\cite{KTW89}, where the
nucleation of droplets with size $R$ of a new state within an old one was
argued to be the main dynamical processes. The free-energy gain of a droplet
formation is caused by the entropic gain due to the exploration of new
states, i.e. by $Ts_{c}R^{3}$, whereas one has to take into account that
such a droplet implies a finite surface energy, characterized by a scale
dependent surface tension, $\sigma \left( R\right) $. A renormalization
group calculation, based on Ref.\cite{Villain84}, leads to the size
dependent surface tension 
\begin{equation}
\sigma (R)=\sigma _{0}\left( R\Lambda \right) ^{-\theta }
\end{equation}
with $\theta =\frac{d-2}{2}$ reflecting the fact that the interface between
two states is wetted by intermediate states. This analysis leads to an
characteristic energy barrier $\Delta E\propto \left( Ts_{c}(T)\right) ^{-1}$
which implies a characteristic relaxation time which follows a Vogel-Fulcher
law 
\begin{equation}
\tau \propto \exp \left( \frac{DT_{K}}{T-T_{K}}\right) ,
\end{equation}
independent on the dimension. Here, fragility parameter \ of the
Vogel-Fulcher law is given by 
\begin{equation}
D=\frac{3\sigma _{0}^{2}}{T^{2}T_{K}\left. \frac{dS_{c}}{dT}\right| _{T_{K}}}%
.  \label{D_frag}
\end{equation}
Xia and Wolynes\cite{XW00} have shown that this scenario gives a
quantitative description of viscous flow in molecular liquids A
straightforward extension of Ref.\cite{KTW89,XW00} along the lines of Ref. 
\cite{XW01} also gives a width of the distributions of different activation
energies characterized by the mean square width $\left\langle \delta
R^{2}\right\rangle $ of the droplet size, which might be compared with the
results of Ref. \cite{CH99}. This droplet picture implies that the glass
breaks up into domains of different metastable states, separated by wetted
surfaces (consisting of intermediate states) leading to a rather small
surface tension. This physical picture is very similar to \ what is usually
called a ''cluster spin glass'', motivated by the observations made in Ref. 
\cite{CBJ92} based on NMR experiments.

\begin{figure}
\centerline{\epsfig{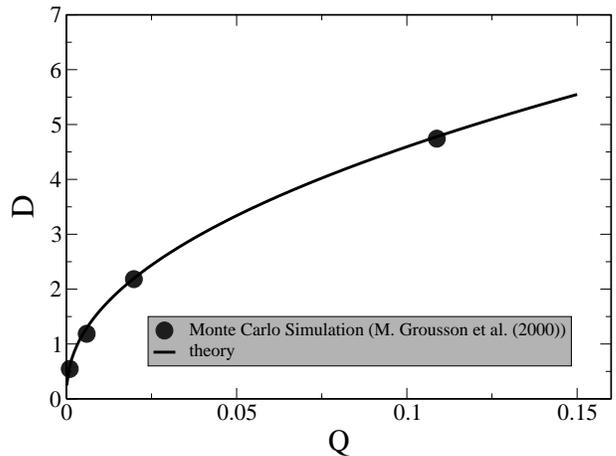}} 
\caption{Comparison of $D$ with
the results obtained in Monte Carlo simulations of the lattice version of
Eq.\ref{ham11} by Grousson {\em et al.}$^{40}$}
\label{fig6}
\end{figure}

For a quantitative analysis we have to develop a theory for the bare surface
tension $\sigma _{0}$ for the stripe glass. This differs from the theory of
molecular liquids because of the long range forces in the model. In what
follows we give simple estimates for $\sigma _{0}$, based on a variational
argument. There are two sources of the bare surface tension in Eq.\ref{ham11}%
, the gradient term and the long range Coulomb term. Assuming a droplet
configuration with locally ordered charge configuration 
\begin{equation}
\varphi \left( r\right) =\varphi _{0}\cos \left( 2\pi r/l_{m}\right) \tanh
\left( \frac{r-R}{l_{w}}\right)
\end{equation}
with droplet radius, $R$, and wall thickness, $l_{w}$. In order to make
progress, we assume \ that the thin wall limit, $l_{w}\ll R$ applies and
find for the gradient term:

\begin{equation}
E_{{\rm s}}^{(1)}=4\pi \varphi
_{0}^{2}l_{w}^{-2}\int_{R-l_{w}/2}^{R+l_{w}/2}r^{2}dr\simeq 4\pi \varphi
_{0}^{2}l_{w}^{-1}R^{2}\ \text{.}
\end{equation}
and for the long range Coulomb term $\ $ 
\begin{equation}
E_{{\rm s}}^{(2)}\simeq 4\pi \varphi _{0}^{2}l_{m}^{-1}R^{2}.\ 
\end{equation}
For larger values of the frustration parameter, $Q$, $l_{w}$ is estimated by
the largest length scale of the problem (except $R$ of course) which should
correspond to the lowest energy of the droplet wall. This gives $l_{w}=\xi $%
. Alternatively, for smaller values of $Q$ the surface tension is dominated
by the contribution of the Coulomb term and we find

\[
\sigma _{0}=4\pi \varphi _{0}^{2}l_{m}^{-1}. 
\]
>From the same variational argument we find $\varphi _{0}^{2}=\frac{1}{u}%
Q^{1/2}$, which gives, together with $l_{m}\simeq 2\pi Q^{1/4}$, the result $%
\sigma _{0}\simeq \frac{2}{u}Q^{3/4}$. Thus, if the frustration parameter
decreases, the barrier height between different metastable states
disappears. Even though the number of metastable states an thus $\left. 
\frac{dS_{c}}{dT}\right| _{T_{K}}$ decreases for decreasing $Q$, the surface
tension term dominates and it follows that 
\begin{equation}
D\left( Q\rightarrow 0\right) \rightarrow 0.
\end{equation}
The fragility parameter does not vanish according to a power law, mostly
because $T_{A}-T_{K}$, which enters $\left. \frac{dS_{c}}{dT}\right|
_{T_{K}}\simeq \frac{S_{c}\left( T_{A}\right) }{T_{A}-T_{K}}$ has
logarithmic behavior as $Q\rightarrow 0$. Essentially, $D$ vanishes with $Q$
similar to a square root. In Fig.\ref{fig6} we compare $D\left( Q\right) $
with the results obtained in Monte Carlo simulations of the lattice version
of Eq.\ref{ham11} by Grousson {\em et al.}\cite{GT01}. Here we multiplied
the result from Eq. \ref{D_frag} by an overall prefactor, leading to a very
good agreement between our analytical theory and the numerical results of 
\cite{GT01}. Since the calculations in Ref. \cite{GT01} are performed for a
lattice version of Eq.\ref{ham11} the actual $T$ dependence of the
correlation length differs from ours. Correspondingly the absolute
magnitudes of $T_{K}$, which enters in Eq. \ref{D_frag}, are different. For
this reason we do not expect the absolute magnitude of $D$, but solely its $%
Q $ dependence to agree in both approaches. 

\section{conclusions}

Glasses are typical examples of systems of many interacting particles that
have a tendency to self-organize into mesoscopic structures. In this paper
we studied the slow activated dynamics of charge inhomogeneities in doped
Mott insulators. We developed an analytical approach which enabled us to
identify the underlying physical mechanism for glassiness in a uniformly
frustrated system. We showed that when the charge correlations are
sufficiently strong, specifically, if $\frac{\xi }{l_{m}}>2$, the stripe
liquid - stripe solid transition can become kinematically inaccessible
because the system undergoes a glass transition, driven by the emergence of
an extensive configurational entropy. We demonstrated that at this point a
Lindemann length, $\lambda $, emerges, which is a length scale over which
imperfections of the stripe pattern manage to wander. Finally, we apply our
results to the scenario of Ref. \cite{KTW89} to calculate the characteristic
relaxation time of the nonequilibrium state. We concluded that the charge
fluctuations in doped Mott insulators have a tendency to self organize into
droplets of metastable states, distinguished by the packing of stripes with
different orientation and the arrangement of defects. These droplets relax
according to a Vogel-Fulcher law, characteristic of structural glasses. We
further compare our results with the doping dependence of $T_{K}$, as
deduced from experiment in Ref. \cite{MCPC00}, and show that it is properly
described within our theory.

\section{ Acknowledgments}

We are grateful to N. Curro, P.C. Hammel, S. A. Kivelson, Z. Nussinov, D.
Pines and G. Tarjus for useful discutions and G. Tarjus for communicating
the results of their Monte Carlo simulations. This research was supported by
an award from Research Corporation (J.S.), the Institute for Complex
Adaptive Matter, the Ames Laboratory, operated for the U.S. Department of
Energy by Iowa State University under Contract No. W-7405-Eng-82 (H.W.Jr.
and J. S.), and the National Science Foundation grant CHE-9530680 (P. G.
W.). H.W.Jr. also acknowledges support from FAPESP.

\bigskip

\appendix

\section{self consistent screening approximation}

In this appendix we summarize the self consistent screening approximation
which was the basis of the numerical investigation of stripe glasses in Ref. 
\cite{SW00} and which is the framework in which we determined the diagonal
and off diagonal element in replica space of the self energy, leading in
particular to Eq.\ref{sigf_calc}. The details of the calculation of Eq.\ref
{sigf_calc} are then given in appendix B.

Eq.\ref{repl_part} has a formal similarity to the action of the random field
Ising model, obtained within the conventional replica approach, which allows
us to use techniques, developed for this model\cite{MY92}. Introducing an $N$%
-component version of Eq.\ref{ham11} with field ${\bf \varphi }=\left(
\varphi _{1},...,\varphi _{N}\right) $ and coupling constant, $u=\frac{u_{0}%
}{N}$, with fixed $u_{0}$ we use a self consistent screening approximation 
\cite{Bray74}, which is exact up to order $1/N$. At the end we perform the
limit $N=1$. The applicability of this approximation is supported by the
strong indications for a non-singular\ large-$N$ limit of Eq.\ref{ham11}, as
discussed in Ref.\cite{NRK99}.

Before we discuss the self consistent screening approximation, it is useful
to summarize a few properties of matrices in replica space with structure
similar to \ref{repans}. Introducing a matrix ${\bf E}$ such that ${\bf E}%
_{ab}=1$ and the unit matrix ${\bf 1}$, it is easy to see that the product
of any two $m\times m$ matrices with structure 
\begin{equation}
{\bf A}=a_{1}{\bf 1}+a_{2}{\bf E,}
\end{equation}
is given by 
\[
{\bf AB}=\left( a_{1}b_{1}\right) {\bf 1}+\left(
a_{1}b_{2}+a_{2}b_{1}+ma_{2}b_{2}\right) {\bf E.} 
\]
\qquad This leads to 
\begin{equation}
{\bf A}^{-1}=\frac{1}{a_{1}}{\bf 1}-\frac{a_{2}}{a_{1}\left(
a_{1}+ma_{2}\right) }{\bf E}  \label{inv_rule}
\end{equation}
for the inverse of \ ${\bf A}$. This \ property was used for the derivation
of Eqs. \ref{Gdyson} and \ref{Dys_od} and will be used below. 
\begin{figure}
\centerline{\epsfig{file=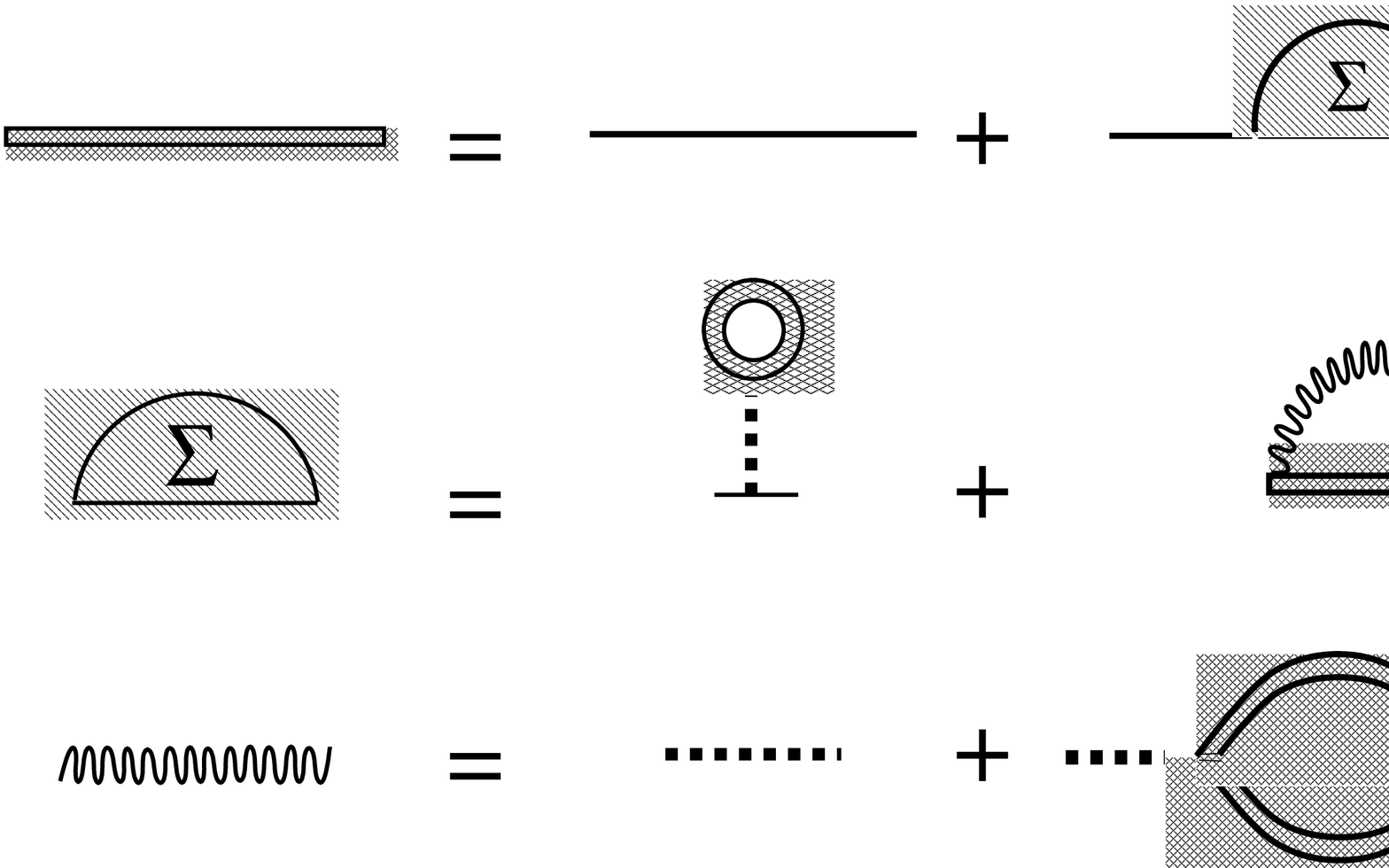,width=8cm,height=6cm,scale=0.85}} 
\caption{Diagrams for the self consistent screening approximation}
\label{fig7}
\end{figure}
The self consistent screening approximation is described by the set of
Feynman diagrams shown in Fig.\ref{fig7}. The self energy is given as 
\begin{equation}
\Sigma _{ab}\left( {\bf q}\right) =\frac{2}{N}\int \frac{d^{3}p}{\left( 2\pi
\right) ^{3}}\ {\cal D}_{ab}\left( {\bf p}\right) {\cal G}_{ab}\left( {\bf %
p+q}\right) .  \label{sigma_app}
\end{equation}
where 
\begin{equation}
{\cal D}\left( {\bf p}\right) =\left( v_{0}^{-1}+\Pi \left( {\bf p}\right)
\right) ^{-1}  \label{D_app}
\end{equation}
is determined self consistently by the polarization function 
\begin{equation}
\Pi _{ab}\left( {\bf p}\right) =\int \frac{d^{3}q}{\left( 2\pi \right) ^{3}}%
{\cal G}_{ab}\left( {\bf q+p}\right) {\cal G}_{ba}\left( {\bf q}\right) .
\end{equation}
In the above set of equations the $p$-integration has to be cut-off at $%
\left| {\bf p}\right| =\Lambda $ and the temperature, $T$, and the coupling
constant, $u_{0}$, occur only in the combination $v_{0}=u_{0}T$. The Ansatz 
\ref{repans} for the Green's function implies an analogous structure for \ $%
\Sigma _{ab}\left( {\bf q}\right) $ and $\Pi _{ab}\left( {\bf q}\right) $ in
replica space. Inserting this ansatz into $\Pi _{ab}\left( {\bf p}\right) $
gives 
\begin{equation}
\Pi =\left( \Pi _{{\cal G}}-\Pi _{{\cal F}}\right) {\bf 1}+\Pi _{{\cal F}}%
{\bf E}
\end{equation}
where the diagonal and off-diagonal elements of the polarization function
are 
\begin{eqnarray}
\Pi _{{\cal G}}\left( {\bf p}\right) &=&\int \frac{d^{3}q}{\left( 2\pi
\right) ^{3}}{\cal G}\left( {\bf q+p}\right) {\cal G}\left( {\bf q}\right) 
\nonumber \\
\Pi _{{\cal F}}\left( {\bf p}\right) &=&\int \frac{d^{3}q}{\left( 2\pi
\right) ^{3}}{\cal F}\left( {\bf q+p}\right) {\cal F}\left( {\bf q}\right) .
\end{eqnarray}

Using the rule \ref{inv_rule}, it is now straightforward to determine ${\cal %
D}_{ab}\left( {\bf p}\right) $ which leads, in the limit $m\rightarrow 1,$
to 
\begin{equation}
{\cal D}=\left( {\cal D}_{{\cal G}}-{\cal D}_{{\cal F}}\right) {\bf 1}+{\cal %
D}_{{\cal F}}{\bf E}
\end{equation}
where 
\begin{equation}
{\cal D}_{{\cal G}}\left( {\bf p}\right) =\left( v_{0}^{-1}+\Pi _{{\cal G}%
}\left( {\bf p}\right) \right) ^{-1}
\end{equation}
and 
\begin{equation}
{\cal D}_{{\cal F}}\left( {\bf p}\right) =-\frac{\Pi _{{\cal F}}\left( {\bf p%
}\right) {\cal D}_{{\cal G}}^{2}\left( {\bf p}\right) }{1-\Pi _{{\cal F}%
}\left( {\bf p}\right) {\cal D}_{{\cal G}}\left( {\bf p}\right) }\ .
\end{equation}

Analogously, inserting the above equations into \ref{sigma_app} we get for
the self energies: 
\[
\Sigma =\left( \Sigma _{{\cal G}}-\Sigma _{{\cal F}}\right) {\bf 1}+\Sigma _{%
{\cal F}}{\bf E} 
\]
where 
\begin{equation}
\Sigma _{{\cal G}}\left( {\bf q}\right) =\frac{2}{N}\int \frac{d^{3}p}{%
\left( 2\pi \right) ^{3}}{\cal D}_{{\cal G}}\left( {\bf p}\right) {\cal G}%
\left( {\bf p+q}\right)
\end{equation}
and 
\[
\Sigma _{{\cal F}}\left( {\bf q}\right) =\frac{2}{N}\int \frac{d^{3}p}{%
\left( 2\pi \right) ^{3}}{\cal D}_{{\cal F}}\left( {\bf p}\right) {\cal F}%
\left( {\bf p+q}\right) . 
\]

The set of equations is closed by the Dyson equation, \ref{dyson} 
\begin{equation}
\left. {\cal G}^{-1}\left( {\bf q}\right) \right| _{ab}=\left( {\cal G}%
_{0}^{-1}\left( {\bf q}\right) +\Sigma _{{\cal G}}-\Sigma _{{\cal F}}\right) 
{\bf 1}+\Sigma _{{\cal F}}{\bf E}.
\end{equation}
that gives, according to \ref{inv_rule}, in the limit $m\rightarrow 1$ the
equations \ref{Gdyson} and \ref{Dys_od}.

In Ref.\cite{SW00} it was shown that this coupled set of equations gives $%
\Sigma _{{\cal F}}\left( {\bf q}\right) \neq 0$ below a characteristic
temperature which was related to the occurrence of glassiness. In the next
appendix we present an approximate analytical solution of this problem.

\section{Analysis of the self energy}

In this appendix we give the main technical details for the calculation of
the off diagonal self energy in replica space, $\Sigma _{{\cal F}}$. The
diagonal self energy $\Sigma _{{\cal G}}$, which turns out to be negligibly
small compared to the leading mean field terms, can be determined in a very
similar fashion and was already analyzed in Ref.\cite{NRK99,ZN00} $\Sigma _{%
{\cal F}}$ and $\Sigma _{{\cal G}}$ were calculated numerically within the
self consistent screening approximation in Ref.\cite{SW00}. \ The virtue of
the analytical calculation presented here is that it is much more
transparent. In all steps of this calculations we did check the reliability
of our approximate analytical treatment by comparing it with the numerical
results.

We start by calculating the polarization function $\Pi _{{\cal G}}(q)$: 
\begin{eqnarray*}
\Pi _{G}\left( q\right) &=&\ \int \frac{d^{3}p}{8\pi ^{3}}{\cal G}_{{\bf p}}%
{\cal G}_{{\bf p+q}} \\
&=&\left\{ \tan ^{-1}\left( \frac{q}{\varepsilon q_{m}}\right) +\frac{1}{2}%
\tan ^{-1}\left( \frac{2q_{m}-q\ }{\varepsilon q_{m}}\right) \right. \\
&&\left. -\frac{1}{2}\tan ^{-1}\left( \ \frac{2q_{m}+q\ }{\varepsilon q_{m}}%
\ \right) \right\} \left( 8\pi q\varepsilon ^{2}\right) ^{-1} \\
&\simeq &\left\{ 
\begin{array}{cc}
\frac{1}{8\pi q_{m}\varepsilon ^{3}} & q<\frac{2\varepsilon q_{m}}{\pi } \\ 
\frac{\Theta (2q_{m}-q)}{16q\varepsilon ^{2}} & q>\frac{2\varepsilon q_{m}}{%
\pi }
\end{array}
\right. .
\end{eqnarray*}
where we used the approximate expression, Eq.\ref{G_app}, for the
correlation function ${\cal G}\left( {\bf x}\right) $. An analogous
calculation for the off diagonal polarization function $\Pi _{{\cal F}}(q)$
gives 
\[
\int \frac{d^{3}p}{8\pi ^{3}}{\cal K}\left( {\bf p}\right) {\cal K}\left( 
{\bf p+q}\right) \simeq \left\{ 
\begin{array}{cc}
\ \frac{1}{8\pi q_{m}\kappa ^{3}} & q<\frac{2\kappa q_{m}}{\pi } \\ 
\frac{\Theta (2q_{m}-q)}{16\kappa ^{2}q} & q>\frac{2\kappa q_{m}}{\pi }
\end{array}
\right. 
\]
as well as 
\[
\int \frac{d^{3}p}{8\pi ^{3}}{\cal G}\left( {\bf p}\right) {\cal K}\left( 
{\bf p+q}\right) \ \simeq \left\{ 
\begin{array}{cc}
\ \frac{1}{8\pi q_{m}\kappa ^{3}} & q<\frac{\kappa +\varepsilon }{\pi }q_{m}
\\ 
\frac{\Theta (2q_{m}-q)}{16\kappa \varepsilon q} & q>\ \frac{\kappa
+\varepsilon }{\pi }q_{m}
\end{array}
\right. . 
\]
For $\left( \varepsilon ,\kappa \right) q_{m}<q<2q_{m}$, we can use the
approximate expressions 
\begin{eqnarray*}
\Pi _{{\cal G}}(q) &=&\frac{1}{16q\varepsilon ^{2}} \\
\Pi _{{\cal F}}(q) &=&\frac{1}{16q}\left( \frac{1}{\varepsilon }-\frac{1}{%
\kappa }\right) ^{2}
\end{eqnarray*}
which gives 
\[
D_{{\cal G}}(q)=\frac{16q_{m}a}{1+\frac{q_{m}}{q}\frac{a}{\varepsilon ^{2}}} 
\]
with dimensionless number $a=\frac{v_{0}}{16q_{m}}\lesssim 1$. $a$ vanishes
as $T\rightarrow 0$, but it always holds that $\varepsilon ^{2}\ll a$. Note,
for the numerical solution in Ref. it holds $a\ \simeq 0.4$. \ Combining
these results and using the fact that $\varepsilon ^{2}/a\ll 1$, the product 
$D_{{\cal G}}(q)\Pi _{{\cal F}}(q)$ becomes momentum independent:

\[
D_{{\cal G}}(q)\Pi _{{\cal F}}(q)\simeq \left( 1-\frac{\varepsilon }{\kappa }%
\right) ^{2} 
\]
and, as a result, $D_{{\cal F}}(q)$ becomes proportional (by a factor
smaller than $1$ in magnitudes) to $D_{{\cal G}}(q)$ 
\begin{eqnarray*}
D_{{\cal F}}(q) &=&\left[ \frac{-D_{{\cal G}}\left( q\right) \Pi _{{\cal F}%
}\left( q\right) }{1-D_{{\cal G}}\left( q\right) \Pi _{{\cal F}}\left(
q\right) }\right] D_{{\cal G}}\left( q\right) \\
&\simeq &\left[ \frac{-\left( 1-\frac{\varepsilon }{\kappa }\right) ^{2}}{%
1-\left( 1-\frac{\varepsilon }{\kappa }\right) ^{2}}\right] D_{{\cal G}}(q)
\end{eqnarray*}

We are now in the position to analyze the self energy $\Sigma _{{\cal F}}$: 
\begin{equation}
\Sigma _{{\cal F}}\left( {\bf q}\right) =\int \frac{d^{3}p}{8\pi ^{3}}D_{%
{\cal F}}({\bf q+p}){\cal F}\left( {\bf p}\right) .
\end{equation}
Since $D_{{\cal F}}\left( {\bf q}\right) $ is only weakly momentum dependent
the same holds for $\Sigma _{{\cal F}}\left( {\bf q}\right) $ and we can
estimate it at the modulation vector. It follows with $t=\sqrt{2\left(
1+\cos \theta \right) }$ 
\begin{eqnarray}
\Sigma _{{\cal F}}\left( q_{m}\right) &\simeq &\int_{0}^{2}tdt\ D_{{\cal F}%
}\left( q_{m}t\right) \int \frac{p^{2}dp}{4\pi ^{2}}{\cal F}\left( p\right) 
\nonumber \\
&\simeq &D_{{\cal F}}\left( q_{m}\right) \int \frac{p^{2}dp}{4\pi ^{2}}{\cal %
F}\left( p\right)
\end{eqnarray}
Using ${\cal F}\left( p\right) ={\cal G}\left( p\right) -{\cal K}\left(
p\right) $ and $\int \frac{p^{2}dp}{4\pi ^{2}}{\cal G}\left( p\right) =\frac{%
1}{8\pi }\left( \frac{q_{m}}{\varepsilon }+\Lambda \right) $ as well as $%
\int \frac{p^{2}dp}{4\pi ^{2}}{\cal K}\left( p\right) =\frac{1}{8\pi }\left( 
\frac{q_{m}}{\kappa }+\Lambda \right) $, we find 
\begin{equation}
\Sigma _{{\cal F}}\left( q_{m}\right) =-\frac{8q_{m}^{2}\varepsilon ^{2}}{%
\pi }\frac{\left( 1-\frac{\varepsilon }{\kappa }\right) ^{2}}{1-\left( 1-%
\frac{\varepsilon }{\kappa }\right) ^{2}}\left( \frac{1}{\varepsilon }-\frac{%
1}{\kappa }\right) .
\end{equation}
Thus, we have determined the off diagonal self energy at the modulation wave
vector as function of the three essential length scales of the problem: $%
l_{m}=\frac{2\pi }{q_{m}}$, $\xi =\frac{2}{\varepsilon q_{m}}$, and $\lambda
=\frac{2}{\sqrt{\kappa ^{2}-\varepsilon ^{2}}q_{m}}$. Note, the dependence
on the momentum cut off, $\Lambda $, cancels completely making this result
robust against lattice corrections.

\section{comparison with the frustration limited domain scenario}

The possibility of glass formation of the model, Eq.\ref{ham11}, has been
pointed out in Ref. \cite{KKZNT95}. As discussed in this appendix, we
disagree with the detailed argumentation of Ref.\cite{KKZNT95}. However the
recognition that a model of the kind of Eq.\ref{ham11} can potentially
describe glass formation was a very important observation.

The aim of Ref.\cite{KKZNT95} was to present an alternative scenario for
glassiness in structural glasses formed of undercooled molecular liquids.
Though the microscopic justification of Eq.\ref{ham11} for the description
of structural glasses is at the least unresolved, one can yet, in principle,
imagine that such long range interactions are caused by a scenario like that
based on icosahedral order which is frustrated by the lack of Euclidean
curvature of the effective space.\cite{DRN85} In what follows we will solely
consider Eq.\ref{ham11} as a given model and leave aside whether it applies
to stripe glasses in doped Mott insulators (as we claim) or to molecular
liquids (as claimed in Ref. \cite{KKZNT95}).

The main idea of Ref.\cite{KKZNT95} is that due to the frustrating
interaction of Eq.\ref{ham11} the system is broken up into ordered domains
of size $R_{D}\sim \xi ^{-1}Q^{-1/2}$ with $\xi $ being the correlation
length that controls the fluctuations inside ordered domains. Furthermore, $%
\xi \ll R_{D}$ is assumed. Within each domain the ordering essentially
corresponds to the one of a finite $Q=0$ system (not of a system with
stripes as in our entropic droplet picture). Since for $Q=0$, $\xi $
diverges at the critical point $T_{c}^{0}$, the avoided frustration scenario
suggests that for $Q>0$, $\xi \left( T\right) $ still has a peak close to $%
T_{c}^{0}$ only rounded due to the frustration. It is then argued that there
is a relaxation rate $\tau ^{-1}$ due to the reorganization of domains
which\ obeys an Arrhenius form 
\begin{equation}
\tau ^{-1}\propto e^{-\Delta F\left( R_{D}\right) /T},
\end{equation}
where $\Delta F\left( R_{D}\right) \sim T_{c}^{0}\left( \frac{R_{D}}{\xi }%
\right) ^{2}$ is the activation energy of the domain. Finally, it is
asserted that the divergence in the viscosity of the system at low
temperatures is determined by $\eta \propto \tau $.

We disagree with this picture. As shown below, the scale $R_{D}$, which
signals the relevance of the long range interaction, can easily be
identified as the Debye screening length of charged particles of size $\xi $
with the expected charge density. Even when the Debye picture applies, we
find it hard to understand how conventional screening can lead to a dynamics
which is dominated by the activated reorganization of screening clouds and
where the natural Langevin description gives a fast relaxation. When $\xi
\ll R_{D}$ there are many short wavelength excitations of weakly coupled
charges and the system should rather behave as a high temperature plasma
than as a glass. In addition, for $N\rightarrow \infty $ it was shown in
Ref. \cite{NRK99} that Debye screening occurs only at temperatures $%
T>T_{c}^{0}$. At low temperatures modulated structures result which lead to
the formation of well correlated stripes as $\xi >l_{m}$. $\xi \left(
T\right) $ does not exhibit a maximum at a temperature comparable to $%
T_{c}^{0}$ but keeps growing until either a stripe solid or a stripe glass
is formed.

We now show that the scale $R_{D}\sim \xi ^{-1}Q^{-1/2}$ is the Debye
screening length of charged particles of size $\xi $. We perform a coarse
graining of the system into regions of linear size $\xi $ centered around
positions ${\bf X}_{i}$. Then Eq.\ref{ham11} becomes: 
\begin{equation}
{\cal H}=\sum_{i}{\cal H}_{i}+\sum_{i>j}\frac{q_{i}q_{j}}{\left| {\bf X}_{i}-%
{\bf X}_{j}\right| }  \label{grain}
\end{equation}
where 
\begin{equation}
{\cal H}_{i}\sim \frac{1}{2}\int_{\xi ^{3}}d^{3}x\left\{ r_{0}\varphi ({\bf %
x)}^{2}+\left( \nabla \varphi ({\bf x)}\right) ^{2}+\frac{u}{2}\varphi ({\bf %
x)}^{4}\right\}
\end{equation}
and charges 
\[
q_{i}\simeq \sqrt{\frac{Q}{8\pi }}\int_{\xi }d^{3}x\varphi ({\bf x)}%
\varpropto \sqrt{Q}\xi ^{5/2}. 
\]
In the last step we assumed that within the volume $\varpropto \xi ^{3}$ the
system is essentially ordered and used $\varphi \varpropto \xi ^{-\beta /\nu
}$, with the critical exponents $\beta =\frac{1}{2}$ and $\nu =\left(
d-2\right) ^{-1}$, obtained within the large $N$ approximation, for $d=3$. 
\cite{Ma76}

The usual analysis of Eq.\ref{grain} within the Debye approximation, i.e.
solving the Poisson equation with induced charges distributed with Boltzmann
weight, gives the Debye screening length: 
\[
l_{D}^{-2}=\frac{2\pi q_{i}^{2}}{T}n, 
\]
where $n$ is the density of charges, here given by $n\simeq \xi ^{-3}$. This
finally gives 
\[
l_{D}\sim Q^{-1/2}\xi ^{-1}, 
\]
which is precisely the length $R_{D}$ proposed on Ref. \cite{NRK99}.

In summary, \ it is unclear why Debye screening should cause a break up of
the system into ordered domains of size $R_{D}=l_{D}$ (remember $\xi \ll
l_{D}$) \ and a slow activated dynamics of such domains, necessary to obtain
large viscosities. Second, for the scenario of Ref. \cite{KKZNT95} to work,
one also has to assume that the correlation length decreases at low
temperature and that Debye theory still applies. The actual analysis of Eq.%
\ref{ham11} for $N\rightarrow \infty $ does not show both assumptions to be
justified\cite{NRK99}. However, it is interesting that for $N<\infty $ the
emergence of a Josephson length scale at low temperatures\cite{NRK99} might
give rise to a competition of physics on two distinct length scales which
then, in principle, could lead to a new long time relaxation along the lines
of the frustration limited domain approach.\cite{SK01} It is interesting to explore the
relationship of this scenario to our replica based theory.

Finally, it is worth pointing out that the entropic droplets discussed in
this paper are qualitatively different from the domains introduced in the
approach of Ref. \cite{KKZNT95}. Whereas the latter correspond to
thermodynamically stable configurations, similar to domains in ferromagnets
caused by the long ranged dipole-dipole interaction, our entropic droplet
are formed by the various metastable states. Transitions between different
droplets are caused by a gain \ of entropy of a system in what would be an
otherwise frozen nonequilibrium state.

\end{multicols}

\end{document}